\documentclass[prx,twocolumn,superscriptaddress,aps,10pt, longbibliography=false]{revtex4-2}
\usepackage[utf8]{inputenc}
\usepackage[american]{babel}
\usepackage[T1]{fontenc}
\usepackage[pdftex]{graphicx}
\usepackage[usenames,dvipsnames,svgnames]{xcolor}
\usepackage{dcolumn}
\usepackage{comment}
\usepackage{scalerel}
\usepackage{lipsum}
\usepackage{footnote}
\usepackage{amsmath}
\usepackage{amsfonts}
\usepackage{amssymb}

\usepackage{cancel}

\usepackage{color}
\usepackage[colorlinks,bookmarks=false,citecolor=NavyBlue,linkcolor=Red,urlcolor=blue]{hyperref}
\usepackage[export]{adjustbox}
\usepackage{amsthm}
\usepackage{simplewick}
\usepackage{bbold}
\usepackage{MnSymbol}
\usepackage{physics}
\usepackage{wasysym}
\usepackage[toc]{appendix}
\usepackage{mathrsfs}
\usepackage{braket}
\usepackage{graphics}
\usepackage{todonotes}
\usepackage{multirow}
\usepackage{orcidlink}

\usepackage{algorithm}
\usepackage{algpseudocode}
\usepackage{dsfont}
\usepackage{mdframed}
\usepackage{tikz,ifthen}
\usepackage{bbold}
\usepackage{tikz-network}
\usepackage{lipsum}
\usetikzlibrary{patterns,decorations.pathreplacing,calligraphy}
\usetikzlibrary{shapes,arrows.meta,decorations.pathmorphing}

\newcommand{\Haar}{{\text{Haar}}}    
\newcommand{\Noise}{{\mathcal{N}}}   

\newcommand{\Wg}{{\text{Wg}}} 

\newcommand{\Id}{\mathbb{1}}
\newcommand{\PauliX}{{\text{X}}}

\newcommand{\Ex}{\mathbb{E}} 

\newcommand{\idp}{e} 
\newcommand{\swp}{s} 

\newcommand{\pnoise}{\noisestrength_{\rm glob}}
\newcommand{\xeb}{\text{XEB}}         
\newcommand{\Ng}{N_{\text{g}}}        
\newcommand{\fidelity}{F}             
\newcommand{\Nth}{\mathsf{L}}         
\newcommand{\PP}{Q}                   
\def\scaleq{\stackrel{\rm s.l.}{=}}   
\def\epsilonRPM{\epsilon}               
\newcommand{\noisestrength}{\gamma}

\newcommand{\new}[1]{\textcolor{black}{#1}}

\definecolor{mycolor1}{rgb}{0.82,0.82,1.}
\definecolor{mycolor2}{rgb}{0.62,0.62,1.}
\definecolor{mycolorbd}{rgb}{0.0,0.0,1.}

\begin{document}

\newcommand{\titleinfo}{Universality in the Anticoncentration of Noisy Quantum Circuits at Finite Depths}
\title{\titleinfo}

\author{Arman Sauliere~\orcidlink{0009-0004-4065-0320}}
\affiliation{Laboratoire de Physique Th\'eorique et Mod\'elisation, CNRS UMR 8089,
CY Cergy Paris Universit\'e, 95302 Cergy-Pontoise Cedex, France}

\author{Guglielmo Lami~\orcidlink{0000-0002-1778-7263}}
\affiliation{Laboratoire de Physique Th\'eorique et Mod\'elisation, CNRS UMR 8089,
CY Cergy Paris Universit\'e, 95302 Cergy-Pontoise Cedex, France}

\author{Corentin Boyer~\orcidlink{0009-0005-4215-5435}}
\affiliation{Laboratoire de Physique Th\'eorique et Mod\'elisation, CNRS UMR 8089,
CY Cergy Paris Universit\'e, 95302 Cergy-Pontoise Cedex, France}

\author{Jacopo De Nardis~\orcidlink{0000-0001-7877-0329}}
\affiliation{Laboratoire de Physique Th\'eorique et Mod\'elisation, CNRS UMR 8089,
CY Cergy Paris Universit\'e, 95302 Cergy-Pontoise Cedex, France}

\author{Andrea De Luca~\orcidlink{0000-0003-0272-5083}}
\affiliation{Laboratoire de Physique Th\'eorique et Mod\'elisation, CNRS UMR 8089,
CY Cergy Paris Universit\'e, 95302 Cergy-Pontoise Cedex, France}
\affiliation{Laboratoire de Physique de l'\'Ecole Normale Sup\'erieure, ENS, Universit\'e PSL, CNRS, Sorbonne Universit\'e, Universit\'e Paris-Cité, 75005 Paris, France}

\begin{abstract}
We present universal properties of anticoncentration in \new{\emph{weakly}} noisy quantum circuits at finite depth. 
\new{We develop a generic framework for single- and multi-qubit noise channels in the weak-noise limit} and introduce an effective description in terms of a \emph{random matrix product operator} (RMPO). 
Within this weak-noise regime, we show that distinct noise mechanisms act in a quantitatively similar way, yielding a universal distribution of bit-string probabilities that is largely independent of the microscopic noise channel and of the circuit architecture. 
We identify three depth-dependent regimes, each characterized by a distinct scaling of cross-entropy benchmarking ($\xeb$) with rescaled depth. 
In the shallow-depth regime, noise effects are perturbatively small; in the intermediate regime, circuit-induced fluctuations and noise compete on equal footing; and in the deep-depth regime, the output distribution becomes effectively classical, up to corrections that are exponentially small in the noise strength. 
We provide quantitative predictions for anticoncentration in generic finite-depth circuits and benchmark them against numerical simulations, finding excellent agreement even at shallow depths. 
Moreover, we show that, contrary to previous expectations, the late-time value of $\xeb$ provides direct access to the global circuit fidelity, even at large noise strengths. 
Our results are directly applicable to current quantum processors and demonstrate universal behavior beyond the pure random-matrix-theory regime, which only emerges at asymptotically large depths.
\end{abstract}

\maketitle



 \section{Introduction}
 Quantum machines have transformative potential in many scientific fields, but current and near-term devices remain limited by external noise~\cite{Arute_2019,boixo2018characterizing,Morvan2023,Quantinuum2025,Liu2025,Yin2025,Haghshenas2025,Shirizly2024,Smith2025,Rost2025,Ringbauer2025,Jacoby2025}. Although advances reduce noise levels, only fault-tolerant error correction can effectively suppress errors \cite{Cafaro2010,Preskill2018,Terhal2015,Quantinuum2025,Aasen2025,Peham2025}. Meanwhile, the development of methods to characterize noisy quantum circuits and to benchmark their output remains crucial~\cite{PhysRevX.12.021021,chen2018classical,markov2018quantum,huang2020classical,aaronson2016complexity,pednault2019leveraging,Gray2021hyperoptimized,bravyi2018simulation,vincent2021jet}.
A widely used benchmarking technique is \emph{random circuit sampling} (RCS), in which one samples bit-strings $\boldsymbol{x}$ from the output density matrix $\rho$ of a quantum circuit $U$ \cite{Arute_2019,Harper2020,Hohloch2021,Choi2023}. By comparing the observed output statistics with those from an ideal (noise-free) classical simulation of the same circuit, one quantifies the hardware global fidelity. This procedure, known as \emph{cross-entropy benchmarking} (XEB), is experimentally accessible and has attracted significant interest~\cite{Arute_2019,Gao_2024,ware2023sharpphasetransitionlinear,Morvan2023,Liu_2024,MarkDaniel2023}.
Since XEB compares two replicas of the system—one noisy and one ideal—it can be readily analyzed using effective circuit averaging techniques developed for random unitary circuits~\cite{Fisher_2023}. However, it captures only partial information about the \emph{anticoncentration} properties of the output distribution in the computational basis~\cite{Harrow2009,Brandao2016,Hangleiter2018,Dalzell2022,luitz2014participation,luitz2014universal,Tirrito2024,mace2019multifractal,Sauliere_2025,Lami_2025,Christopoulos_2025,Magni2025,Claeys2025,MarkDaniel2023,Daniel2024}.
The latter are instead fully encoded in the so-called \emph{probability-of-probabilities distribution} (PoP) \cite{Kaufman2024,Arute_2019,Shaw_2024} $P(w)$, i.e.\ the average distribution of output probabilities $w = D\langle\boldsymbol{x} | \rho |\boldsymbol{x}\rangle$ (for convenience rescaled by the Hilbert space dimension $D$). The moments of the PoP are proportional to the celebrated \emph{Inverse Participation Ratios} (IPR), often used to signal localization and non-ergodicity effects~\cite{DeLuca_2013, PhysRevLett.123.180601}.

\begin{figure}[h!]
\includegraphics[width=1.01\linewidth]{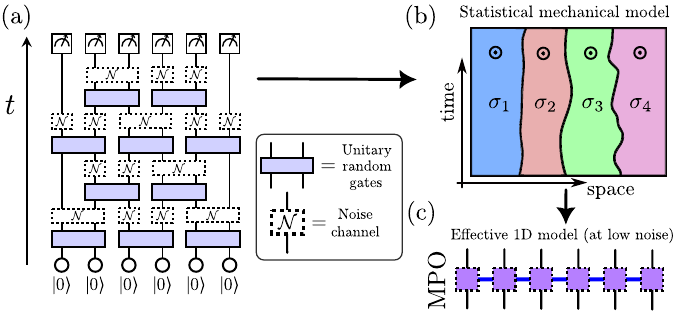}
\caption{Sketch of the work. (a) We consider the task of random circuit sampling. The circuit is composed by random unitary gates in the usual brickwall geometry, and it is affected by noise represented by a quantum channel $\mathcal{N}$. We explore noise acting either on single or two-adjacent qubits, all yielding similar conclusions. (b) Theoretical arguments show that the gate average of the replicated circuit gives rise to an effective one-dimensional statistical model of permutations. In this model, noise acts as an external field that biases the system toward permutations close to the identity. (c) As a result, the original spatio-temporal circuit becomes equivalent to a one-dimensional random Matrix Product Operator (MPO) which incorporate noise.}
\label{fig:1}
\end{figure}

In the absence of noise, the PoP for Haar-random quantum states follows the Porter–Thomas (PT) distribution~\cite{PhysRev.104.483}, that for $D \to \infty$ takes the simple form $P_{\text{PT}}(w)=e^{-w}$. 
The same distribution is obtained for generic circuits of sufficiently large depth in the absence of any conserved quantities. By contrast, generic classical stochastic processes produce uniform bit-string sampling, implying $P_{\text{C}}(w) = \delta(w - 1)$, a distribution that also characterizes quantum devices overwhelmed by noise. In general, a noisy quantum circuit is expected to interpolate between these two extremes as a function of the noise strength. Additionally, these forms hold strictly in the infinite-depth limit~\cite{Kaufman2024,Arute_2019,Shaw_2024}; capturing the quantum–classical crossover 
requires incorporating finite-depth corrections.

\vspace{5 mm} 
\section{Summary of results}

In this work, we analyze the combined impact of finite circuit depth and noise on both the linear cross-entropy benchmarking (XEB) estimator and the full probabilities-of-probabilities (PoP) distribution in a \new{short-range}, $1{+}1$-dimensional quantum circuit with open boundary conditions. \new{We focus on the \emph{weak-noise regime}~\cite{Brandao_2021,ware2023sharpphasetransitionlinear,Dalzell2024,Gao_2024}, where $\eta$, a measure of the total number of errors in the circuit, can be treated as an \emph{intensive} parameter, i.e., it does not grow linearly with the system size. We obtain explicit and exact results in the thermodynamic limit $N\to
\infty$ at fixed $\eta$. It is clear that the most physical assumption in realistic setups is that the number of errors $\eta$ in a circuit is at least proportional to the number of gates and qubits. However, state-of-the-art platforms operate at a very low error rate~
\cite{Morvan2023,Klimov2024,2025googleotoc,Manetsch2025,Chiu2025}, so that for the $N$ currently accessible, one effectively has $\eta \ll N$, a regime in which, as we shall see, our predictions prove useful and accurate.}

We show that, in this weak-noise regime, both the XEB and the PoP distributions exhibit a \emph{universal scaling regime} under \emph{generic forms of noise}. This universal scaling is characterized by two microscopic parameters that can be accessed experimentally from just the XEB: $\eta$, directly related to the global fidelity $F=e^{-\eta}$, and $x$, a rescaled inverse depth defined as
\begin{equation}\label{eq:def_x}
    x = \frac{N}{\Nth(t)},
    \qquad
    \Nth(t) = \Nth_0\, e^{t/\tau}\;.
\end{equation}
The effective length $\Nth(t)$ is the characteristic scale controlling the approach to random-matrix behavior, often referred to in quantum chaos as a \emph{Thouless length}: it marks the scale at which the circuit locally behaves as if it were fully random~\cite{RPMPhysRevLett.121.060601,Shivam_2023,chan2021manybody,Christopoulos_2025,Lami_2025}. The growth of $\Nth(t)$ in time---which can be viewed as a reparameterization of circuit depth---occurs on the same timescale $\tau$ that governs the decay of the half-chain purity. Denoting by $\rho_{N/2}$ the reduced density matrix of half the system and by $P_{N/2}=\Tr(\rho_{N/2}^2)$ the corresponding purity, a standard entanglement diagnostic, we have that, starting from a generic product state, $P_{N/2}\sim e^{-t/\tau} \sim \Nth(t)^{-1}$~\cite{Turkeshi2024,Christopoulos_2025,Sauliere_2025}. The rescaled depth $x$ in~\eqref{eq:def_x} thus targets a shallow-circuit regime, where time scales logarithmically with system size, namely $t=O(\log N)$.

Within this scaling regime, we identify \emph{three dynamical regimes}, determined by the interplay between inverse depth $x$ and noise $\eta$, and we fully characterize the PoP distribution $P_{x,\eta}(w)$ in each of them. This yields a quantitative description of the quantum-to-classical crossover at finite depth, i.e., the progressive loss of distinctly quantum sampling features as noise and finite depth compete.

Our \emph{universal results} rely on the observation that, in the weak-noise limit and within the scaling regime~\eqref{eq:def_x}, generic chaotic circuits subject to different noise mechanisms can be reduced to an effective \emph{random matrix product operator} (RMPO), a tractable random-tensor description of the noisy evolution (see also Fig.~\ref{fig:1}). The RMPO construction extends the notion of \emph{random matrix product states} (RMPS)---random tensor-network models used to capture typical properties of pure states~\cite{garnerone2010typicality,garnerone2010statistical,haag2023typical,lancien2021correlation,PRXQuantum.6.010345,7p1r-r2p6}---to noisy channels. This is achieved by doubling the Hilbert space (a standard purification technique to represent mixed states and noise processes) and by incorporating the noisy evolution step by step along the circuit.

We establish this mapping explicitly in a controlled setting, namely the Random Phase Model (RPM)~\cite{RPMPhysRevLett.121.060601,Christopoulos_2025}, in the limit of large local Hilbert-space dimension. In this limit we obtain explicit expressions for XEB and for the moments of the outcome probabilities, and we introduce an efficient numerical procedure to reconstruct the full PoP distribution. Through extensive numerical simulations, we show that the scaling regime provides a quantitatively accurate description also for random circuits with brick-wall geometry, a standard architecture widely used in many-body quantum chaos studies~\cite{Nahum2018Operator,Fisher_2023,PhysRevLett.132.010401,ChanDeLucaChalker2018,ChanDeLucaChalker2018PRL,ChanDeLucaChalker2019,Bertini2019Exact,Piroli2020Exact,Parker2019Operator,Cotler2017Early,Gopalakrishnan2020Hydrodynamic,Gopalakrishnan2020Hydrodynamic2,HunterJones2018Onset}, and for several experimentally relevant noise models.

Finally, we consider the more realistic situation in which the number of errors grows with circuit depth, $\eta=\lambda t$, \new{while remaining within the regime $\lambda t \ll N$, corresponding either to shallow circuits or sufficiently weak noise, where our results remain applicable}. In this setting, we use our theory to derive a universal late--intermediate-time form for $\xeb$ that captures the crossover in its decay rate as the noise strength is varied~\cite{Morvan2023}. Contrary to what was previously believed, we also show that, even at generic noise strength, $\xeb$ can still be used to extract the circuit's global fidelity.

Overall, our results provide a unified framework for understanding how noise and finite depth jointly shape sampling statistics and benchmarking metrics in near-term quantum processors.

\section{Setup and Methods}

\new{This section introduces the main notation, quantities, and setup
used throughout the paper.} \new{Table~\ref{tab:notation} at the end of this section collects the notation used throughout the rest
of the manuscript (with some symbols defined in subsequent sections).}

\subsection{Setup}
We consider a system of $N$ qudits with local dimension $d$, and total Hilbert space dimension $D = d^N$.
The computational basis is labeled by bit strings $|\boldsymbol{x}\rangle = |x_1, \dots, x_N\rangle$, where $x_i \in \{0, \dots, d-1\}$. The task of \emph{Random Circuit Sampling} (RCS) is to apply a random (but fixed) quantum circuit $U$ to the initial state $\lvert \mathbf{0} \rangle= |0, \dots, 0\rangle$, which yields $\rho(U) \;=\; U \,\lvert \mathbf{0} \rangle\langle \mathbf{0} \rvert \,U^{\dagger}$, and then use quantum measurements to sample bit strings $\boldsymbol{x}$ from the \emph{output distribution} $p(\boldsymbol{x}; \, \rho(U))$, where we define
\begin{equation}\label{eq:overlap0}
    p(\boldsymbol{x};\rho)
    := \langle \boldsymbol{x}| \rho | \boldsymbol{x}\rangle \, .
\end{equation}
To fix ideas, we consider a brickwall circuit with alternating layers of two-qudit gates, all independently drawn from the Haar measure $\Haar(d^2)$ (see Fig.~\ref{fig:1}). \\

\subsection{Noise models}\label{sec:sub_noise}
In realistic settings, gates are subject to noise, which we model by a quantum channel of the form
\begin{equation}
\label{eq:noisechandef}
    \Noise(\rho) = \sum_{\alpha} K_{\alpha}\, \rho\, K_{\alpha}^\dagger \;, \quad \sum_{\alpha} K_{\alpha}^\dagger K_{\alpha} = \Id
\end{equation}
where $K_{\alpha}$ are Kraus operators. Under this noise, the system evolves into a noisy version $\rho_{\Noise}(U)$ of the final state, and consequently the output distribution becomes $p(\boldsymbol{x};\rho_{\Noise}(U))$.
\new{
To discuss the weak noise limit, we assume that $\Noise$ belongs to a smooth family of completely positive and trace-preserving (CPTP) map passing through the identity, parameterized via a formal microscopic positive parameter $\noisestrength$. Consequently, both the channel and its Kraus operators depend explicitly on this parameter, i.e.\ $\Noise \to \Noise_{\noisestrength}$ and $K_{\alpha} \to K_{\alpha}(\noisestrength)$. 
The parameter $\noisestrength$ here represents the dominant scale for noise in the setup quantifying the \textit{strength} of the noise channel $\Noise$. Consequently,  for $\noisestrength = 0$, the noise channel reduces to the identity, namely $\lim_{\noisestrength \to 0^+} \Noise_{\noisestrength}(\rho) = \rho$, and $\lim_{\noisestrength \to 0^+} K_{\alpha}(\noisestrength) = \delta_{\alpha,0} \Id$. 
As we will see explicitly, this parameterization does not restrict the ensemble of possible noise sources we consider, nor the coexistence of different noises with comparable intensities, as long as they all remain weak.
Indeed, in Section~\ref{sec:noisy_wg} we provide an explicit example for a gate depending on two noise parameters.} As an explicit example, the depolarizing noise channel is given by
\begin{equation}\label{eq:depp}
\Noise_{\noisestrength}^{\text{dep.}}(\rho) = (1- \noisestrength)\rho +  \noisestrength \, \frac{\Id}{q} \, ,
\end{equation}
where $\noisestrength \in [0,1]$ and $q$ is the dimension of the Hilbert space on which the noisy channel acts. In general, we consider scenarios where the noise channel acts on single qudits ($q = d$) or on multiple qudits (for example, two qudits with $q = d^2$). \\

\new{
Since we focus on the weak-noise regime, we expand $\Noise_{\noisestrength}$ at first-order in $\gamma$, so that the action of the noisy channel becomes $\Noise_{\noisestrength}(\rho)=\sum_{\alpha} K_\alpha(\noisestrength) \rho K_\alpha^\dag(\noisestrength) = \rho + \noisestrength \tilde{\Noise}(\rho)+ O(\noisestrength^2)$, with $\tilde{\Noise}(\rho)$ a Lindblad operator~\cite{breuer2002theory}.
Explicitly, from the standard first-order expansion of the Kraus operators
\begin{equation}\label{eq:krauss_expansion}
\left\{
\begin{array}{l}
K_0(\noisestrength)=\mathbb{1}-\noisestrength \tilde{K}_0 +O(\noisestrength^2).\\
K_{\alpha}(\noisestrength)=\sqrt{\noisestrength}\tilde{K}_{\alpha} +O(\noisestrength^{3/2})\quad \text{  for   } \,  \, \alpha \geq 1,
\end{array}
\right.
\end{equation}
}
which implies
\begin{equation}\label{eq:noise_tilde}
\tilde{\Noise}(\rho) = \sum_{\alpha \geq 1 } \tilde{K}_\alpha\rho \tilde{K}_\alpha^\dag - \tilde{K}_0 \rho - \rho \tilde{K}_0^{\dag}  \, .
\end{equation}
The trace-preserving condition for the noise channel implies $\tilde{K}_0+\tilde{K}_0^\dag=\sum_{\alpha \geq 1} \tilde{K}_{\alpha}^\dag \tilde{K}_{\alpha}$. The Kraus operators are not uniquely defined, and we fix this freedom by choosing the $\tilde{K}_{\alpha}$ to be traceless. \new{Notice also that one has the freedom to rescale $\gamma$ and the operators $\tilde{K}$. In particular, the transformation $\gamma \to \gamma / a, \quad \tilde{K}_{0} \to a\, \tilde{K}_{0}, \quad \tilde{K}_{\alpha} \to \sqrt{a}\, \tilde{K}_{\alpha}$ (for any value of $a$) leaves Eq.~\eqref{eq:krauss_expansion} unchanged.
In any case, as we anticipated, the expansion to the first order in $\noisestrength$ still leaves great freedom in the choice of an arbitrary Lindblad operator $\tilde\Noise$ and the corresponding jump operators $\tilde{K}_\alpha$}.\\

\subsection{Anticoncentration properties}
A crucial quantity to assess the performance of RCS is the \emph{linear cross-entropy benchmark} (XEB)
\begin{equation}
\label{eq:xebdef}
    \xeb(U) = D \sum_{\pmb{x}} p(\boldsymbol{x};\rho_{\Noise}(U)) \, \, p(\boldsymbol{x};\rho(U)) - 1 \, ,
\end{equation}
which quantifies the correlation between the ideal and the noisy distributions.
For intermediate system sizes, $\xeb$ can be estimated by combining experimental data with a classical simulation that gives direct access to $p(\boldsymbol{x};\rho(U))$~\cite{Arute_2019} and, for deep circuits ($t \gg \log N$, i.e.\ $x \to 0$), it has been used as an experimentally friendly proxy for the fidelity $\fidelity(U) = \Tr[\rho_{\Noise}(U) \rho(U)]$ between the ideal target state and its noisy counterpart; however, recent works have pointed out that this correspondence may only hold when the noise is sufficiently weak \cite{Brandao_2021, Gao_2024, ware2023sharpphasetransitionlinear,Morvan2023}.

Other key quantities we analyze are the PoP and its moments, represented by the \textit{Inverse Participation Ratios} (IPRs)~\cite{Shaw_2024}, defined respectively as
\begin{align}\label{eq:definitions}
  \begin{split}
     P(w; U) &= D^{-1} \sum_{\pmb{x}} \delta( w - D \, p(\pmb{x}; \rho_{\Noise}(U))) \, , \\
    I_{k}(U) &= \sum_{\pmb{x}} p(\pmb{x}; \rho_{\Noise}(U))^{k} \, \, \text{  with  } \, \,  k \geq 1  \, .
   \end{split}
\end{align}
These quantities probe the anticoncentration of the final state in the computational basis. Up to a constant factor, the IPR equals the $k$-th moment of the PoP: $\Ex[w^k] = D^{k-1} I_{k}$.

\subsection{Weingarten calculus}
While the quantities defined above depend explicitly on the specific circuit instance $U$, one can average them over an ensemble of random circuits, thereby defining $\xeb = \Ex_U[\xeb(U)]$, $F=\Ex_U[\fidelity(U)]$, $I_{k} = \Ex_U[I_{k}(U)]$, and $P(w) = \Ex_U[P(w; U)]$. To compute such circuit averages $\Ex_U[\dots]$, we employ Weingarten calculus~\cite{Kostenberger_2021,Collins_2022}, which expresses the Haar average of $k$ copies of a unitary gate as a sum over permutation operators $\sigma \in S_k$, where $S_k$ is the symmetric group of order $k$. Specifically, using the vectorized representation of operators $O \to |O\rangle\!\rangle$ and of their inner product $\mathrm{tr}(O^\dagger Q) = \langle\!\langle O | Q \rangle\!\rangle$, the Weingarten formula reads~\cite{Kostenberger_2021,Collins_2022}
\begin{equation}\label{eq:weingarten}
\mathbb{E}_{U \sim \Haar(q)} \left[ (U^* \otimes U)^{\otimes k} \right] = \sum_{\pi, \sigma \in S_k} \Wg_{\pi, \sigma}(q)\, |\pi\rangle\!\rangle \langle\!\langle \sigma| \, .
\end{equation}
Here, $\Wg(q)$ denotes the $k! \times k!$ Weingarten matrix, which is the pseudo-inverse of the overlap matrix $G_{\pi,\sigma}(q):=\llangle \pi |\sigma\rrangle_q$ (the subscript indicates the Hilbert space dimension).
For a $q$-dimensional Hilbert space, vectorized permutations can be formally expanded in the computational basis as
\begin{equation}\label{eq:sigma_expanded}
   |\sigma\rangle\!\rangle = \sum_{x_1, \dots, x_k = 0}^{q-1} |x_1, x_{\sigma^{-1}(1)} \rangle \dots |x_k, x_{\sigma^{-1}(k)} \rangle \, .
\end{equation}
We can represent these states graphically using standard tensor-network notation. For example, with $k = 4$ replicas, Eq.~\eqref{eq:sigma_expanded} for the permutation $\sigma = (1)(2\ 3\ 4)$ reads
\begin{equation}\label{eq:perm_neww}
\vcenter{\hbox{\includegraphics[width=0.655\linewidth]{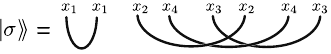}}}
\end{equation}
where we explicitly label the indices represented by the lines.
\new{In the rest of the paper---particularly in Section~\ref{sec:RMPOmodel}---we will make use of the large-$q$ expansion of the overlap and Weingarten matrices to derive explicit universal forms for the IPRs and $\xeb$. This expansion reads}
\begin{align}\label{eq:expanding_GWG}
\begin{split}
G(q) &= q^{k} \left( \Id +  q^{-1} \, A + O(q^{-2}) \right) \\
\Wg(q) &= q^{-k} (\Id - q^{-1} \, A + O(q^{-2})) \, ,
\end{split}
\end{align}
where $A$ is a matrix 
whose coefficients $A_{\pi, \sigma}$ are non-vanishing and equal to one only if $\tau,\sigma$ that differ by a single transposition. 

\subsection{Porter-Thomas distribution}

In the absence of noise and for global random unitaries $U \sim \mathrm{Haar}(D)$---or, equivalently, for circuits of sufficiently large depth---the PoP follows the well-known \emph{Porter-Thomas} (PT) distribution
\begin{equation}\label{eq:PTDistribution}
    P_{\text{PT}}(w) = e^{-w} \, ,
\end{equation}
and correspondingly the IPRs are $I_k^{\text{PT}} = k! D^{1-k}$ (both equalities hold for $D \gg 1$). The absence of noise results in a perfect cross-entropy score $\xeb_{\text{PT}} = D I_2 - 1 \simeq 2 - 1 = 1$.

To illustrate how noise affects these results, we consider the simple scenario of a global depolarizing channel (Eq.~\eqref{eq:depp}) $\rho_{\mathrm{noise}} = \mathcal{N}_{\pnoise}^{\text{dep.}}(U |\pmb{0}\rangle \langle \pmb{0}| U^{\dag})$, with $U \sim \mathrm{Haar}(D)$ as before. In this case, the bitstring distribution is $p(\boldsymbol{x};\rho_{\Noise}(U)) = (1-\pnoise) \, p(\boldsymbol{x};\rho(U)) + \frac{\pnoise}{D}$, and therefore the IPR is $I_{k}(\pnoise) = D \sum_{j=0}^k \binom{k}{j} (1-\pnoise)^{j} (\frac{\pnoise}{D})^{k-j} \,\Ex_U[p(\pmb{0}; \rho(U))^{j}]$. Using the moments of the PT distribution, we have $\Ex_U[p(\pmb{0}; \rho(U))^{j}]=j! D^{-j}$. Moreover, it is useful to expand the factorial as $j!=\sum_{i=0}^j !i \cdot \binom{j}{i}$, where $!i$ denotes the number of derangements of a set of size $i$, i.e., the number of permutations with no fixed points~\cite{brualdi2012introductory}. Exchanging the sums over $j$ and $i$ yields
\begin{equation}\label{eq:noiseIkglobal}
I_k(\pnoise)=D^{1-k} \Tr[ e^{\PP \, \ln(1-\pnoise)}] \, ,
\end{equation}
with
\begin{equation}\label{eq:noiseOp}
\PP_{\sigma' \sigma} = \delta_{\sigma' \sigma} \, (k - \mathrm{n_F}(\sigma))
\end{equation}
a diagonal $k! \times k!$ matrix that depends on the number of fixed points $\mathrm{n_F}(\sigma)$ of the permutation, and plays a central role in our subsequent analysis. \new{As an explicit example, for $k=4$ the permutation $\sigma = (1)(2\,3\,4)$ shown in Eq.~\eqref{eq:perm_neww} has a single fixed point (the element $1$), implying $\mathrm{n_F}(\sigma)=1$ and $\PP_{\sigma\sigma}=3$.} \new{More generally, the weight $e^{\PP \ln(1 - \pnoise)}$ favors permutations with more fixed points, and in the strong-noise limit it suppresses all but the identity $e$ (for which $\PP_{ee} = 0$).}
\new{For any value of $\pnoise$, the PoP associated with the moments in Eq.~\eqref{eq:noiseIkglobal} is a \emph{shifted Porter-Thomas} (SPT) distribution, given by}
\begin{equation}\label{eq:shiftedPT}
    P^{\text{SPT}}_{\pnoise}(w) = (1-\pnoise)^{-1} e^{-\frac{w-\pnoise}{1-\pnoise}} \theta\left( w-\pnoise \right)
\end{equation}
where $\theta(w)$ is the Heaviside function. The noise strength $\pnoise$ is directly related to the XEB score, as $\xeb_{\text{SPT}} = 1 - \pnoise$.
In the strong-noise limit $\pnoise \to 1$, the SPT distribution converges to $P_{\text{C}}(w) = \delta(w - 1)$, with corresponding IPRs $I_k^{\text{C}} = D^{1 - k}$, reflecting the dominance of just the identity permutation $e$ over all others. This delta distribution $P_{\text{C}}(w)$ can thus be interpreted as the purely classical (i.e.\ non-quantum) form of the PoP. \\

\begin{table}[h]
\centering
\begin{tabular}{c c}
\hline
\textbf{Symbol} & \textbf{Meaning} \\
\hline
$N$ & System size (number of qudits) \\
$d$ & Local Hilbert space dimension \\
$D$ & Total Hilbert space dimension \\
$q$ & Generic Hilbert space dimension \\
$t$ & Circuit depth \\
$\tau$ & Decay rate of half-system purity (noiseless) \\
$x$ & Rescaled inverse circuit depth $x=  e^{-t/\tau} \cdot N/\Nth_0$ \\
$\noisestrength$ & Microscopic noise parameter / strength \\
$\lambda$ & Number of errors per circuit layer \\
$\eta$ & Total number of errors, fixed by global fidelity $F=e^{-\eta}$ \\
\hline
\end{tabular}
\caption{Notation used throughout the manuscript.}\label{tab:notation}
\end{table}

\section{Noisy Weingarten calculus and weak-noise expansion}\label{sec:noisy_wg}


\new{To evaluate the IPRs and $\xeb$, one must average over ensembles of
noisy random quantum circuits. This requires a noisy version of the
standard Weingarten formula, Eq.~\eqref{eq:weingarten}. Here, we
explicitly derive it for the first time for a specific noise model (depolarizing noise). We further argue that the \emph{weak-noise expansion} is universal, i.e.\ it
holds \emph{independently of the particular noise model.}}

\subsection{Depolarizing noise}\label{sec:depp}
\new{As a first example, we consider a single unitary gate $U$ subjected to depolarizing noise (Eq.~\eqref{eq:depp}).
In vectorized form, the combined effect of the gate and noise is}
\begin{align}
\begin{split}
&\Noise_{\noisestrength}^{\text{dep.}}\left((U \otimes U^{*})|\rho\rangle\!\rangle\right)= \\
&=\left((1- \noisestrength) \, U \otimes U^{*}  +  \frac{\noisestrength}{q} |\idp \rrangle \llangle \idp| \right)   |\rho\rangle\!\rangle  \, ,
\end{split}
\end{align}
\new{where $|e\rangle\!\rangle$ is the (folded) identity permutation.
To carry out replica calculations, it is necessary to average over $k$ replicas of the noisy gate, i.e.\ to compute
$\Ex_U[\big(  (1- \noisestrength) \, U \otimes U^{*}  +  \frac{\noisestrength}{q} |\idp \rrangle \llangle \idp|  \big)^{\otimes k}]$.
Since the depolarizing channel commutes with unitary conjugation, the average therefore mirrors that of the noiseless case
}
\begin{align}\label{eq:noisy_wg}
    \begin{split}
& \Ex_{U \sim \Haar(q)}[\left(  (1- \noisestrength) \, U \otimes U^{*}  +  \frac{\noisestrength}{q} |\idp \rrangle \llangle \idp|  \right)^{\otimes k}] = \\
&=\sum_{\pi, \sigma \in S_k} \tilde{\Wg}_{\pi, \sigma}^{\text{dep.}}(q,\noisestrength)\, |\pi\rangle\!\rangle \langle\!\langle \sigma| \, ,
    \end{split}
\end{align}
where $\tilde{\Wg}_{\pi, \sigma}^{\text{dep.}}(q,\noisestrength)$ denotes \emph{noisy Weingarten coefficients} that remain to be determined. To find these coefficients, we fix $\sigma$ and $\pi$, and observe that each common fixed point of the two permutations can arise in the expansion of the left-hand side of Eq.~\eqref{eq:noisy_wg} from a term $|\idp \rrangle \llangle \idp|$ inserted at a specific replica, while the remaining part of the permutations must come from the Haar average of $U \otimes U^{*}$. Taking these observations into account, we arrive at
\begin{equation}
    \label{eq:Noisy Weingarten}
    \tilde{\Wg}^{\text{dep.}}_{\pi,\sigma}(q,\noisestrength)=\sum_{i=0}^{\mathrm{n_F}(\pi,\sigma)} \binom{\mathrm{n_F}(\pi,\sigma)}{i}\frac{\noisestrength^i}{q^{i}} (1-\noisestrength)^{k-i}\Wg_{\tilde{\pi}(i),\tilde{\sigma}(i)}(q) \, ,
\end{equation}
where $\mathrm{n_F}(\pi,\sigma)$ denotes the number of common fixed points {of $\sigma \pi^{-1}$}, and $\tilde{\pi}(i)$ and $\tilde{\sigma}(i)$ denote the permutations obtained from $\pi$ and $\sigma$, respectively, by removing $i$ of these common fixed points (which ones is immaterial, because $\Wg$ depends only on the cycle structure). \\

\new{We now consider the weak-noise expansion of the noisy Weingarten coefficients. At first order in $\noisestrength$, Eq.~\eqref{eq:Noisy Weingarten} reads}
\begin{align}\label{eq:noise_wg_exp0}
    \begin{split}
&\tilde{\Wg}^{\text{dep.}}_{\pi,\sigma}(q,\noisestrength)= \\
&= (1-k \noisestrength)\Wg_{\pi,\sigma}(q)
+ \mathrm{n_F}(\pi,\sigma) \frac{\noisestrength}{q} \Wg_{\tilde{\pi}(1),\tilde{\sigma}(1)}(q)
+ O(\noisestrength^2) \, .
    \end{split}
\end{align}
\new{This provides the leading non-trivial correction to the standard Weingarten matrix induced by noise.}
\new{Beyond the weak-noise assumption, in the remainder of the paper---particularly in Section~\ref{sec:RMPOmodel}---we will also consider the regime in which the relevant Hilbert space dimension $q$ is large; in our effective RMPO model, this corresponds to a large-bond-dimension limit. Specifically, we take $q^{-1}$ to be of the same order as $\noisestrength$. Using Eq.~\eqref{eq:expanding_GWG} to expand Eq.~\eqref{eq:noise_wg_exp0} for large $q$ yields}
\begin{align}\label{eq:noise_wg_exp}
    \begin{split}
      &\Tilde{\Wg}^{\text{dep.}}(q,\noisestrength) = \\
      &=q^{-k}\left(1-  \frac{A}{q}-  \noisestrength \, \PP + O(q^{-2}, \noisestrength^{2}, q^{-1} \noisestrength) \right)   \, ,
    \end{split}
\end{align}
\new{with the matrix $\PP$ defined as in Eq.~\eqref{eq:noiseOp}. The expansion in Eq.~\eqref{eq:noise_wg_exp} represents the main result of this section.} \\

\subsection{Generic noise}
\label{subsec:Generic noise}

\new{To show explicitly that the weak-noise expansion of the noisy Weingarten coefficients in Eq.~\eqref{eq:noise_wg_exp} is model independent, we repeat the derivation using a \emph{generic noise channel} $\Noise_{\noisestrength}$. As in the previous case, we consider the averaged action of a gate $U$ followed by noise. However, {as we are interested in the regime of rare noise events}, we assume that the noise is subsequently scrambled by a second gate $V$, with both $U$ and $V$ Haar random. The averaged combined action of the unitary gates and noise on $k$ copies of the system is described by
$\Ex_{U,V \sim \Haar(q)}\!\left[\left( (V \otimes V^{*}) \cdot \Noise_{\noisestrength} \cdot (U \otimes U^{*}) \right)^{\otimes k} \right]$, which we aim to evaluate. Note that in the noiseless case ($\gamma=0$), Haar invariance implies that the product $UV$ is still Haar random, so the expression reduces to Eq.~\eqref{eq:weingarten}. For generic values of $\gamma$, we can apply Eq.~\eqref{eq:weingarten} to average over both $U$ and $V$, obtaining}
\begin{align}\label{eq:noisy_wg_new_1}
    \begin{split}
&\Ex_{V \sim \Haar(q)}[(V \otimes V^{*})^{\otimes k}] \, \, \Noise_{\noisestrength}^{\otimes k} \, \, \Ex_{U \sim \Haar(q)}[(U \otimes U^{*})^{\otimes k}] = \\
&=\sum_{\pi, \sigma \in S_k} \tilde{\Wg}_{\pi, \sigma}(q,\noisestrength)\, |\pi\rangle\!\rangle \langle\!\langle \sigma| \, ,
    \end{split}
\end{align}
\new{where $\tilde{\Wg} = \Wg \, \tilde{G} \, \Wg$ are modified Weingarten coefficients and}
\begin{align}
    \begin{split}
\tilde{G}_{\sigma \pi}(q,\gamma) &= \langle\!\langle \sigma| \Noise_{\noisestrength}^{\otimes k} |\pi\rangle\!\rangle_q
    \end{split}
\end{align}
\new{is a modified version of the standard overlap matrix $G(q)$. In particular, $\tilde{G}$ reduces to $G$ in the noiseless limit $\gamma = 0$: $G(q) = \tilde{G}(q,0)$.}  \\
\new{Technically, note that the action of the noise operator $\Noise_{\noisestrength}^{\otimes k}$ on a permutation state $|\sigma\rangle\!\rangle$ can generate components outside the vector space spanned by permutations. This does not occur for depolarizing noise (Section~\ref{sec:depp}), but it can arise for more general noise models. However, in the present construction, averaging over an additional Haar-random unitary---i.e.\ applying $\mathbb{E}_{V \sim \Haar(q)}\!\left[(V \otimes V^{*})^{\otimes k}\right]$---projects the state back onto the permutation subspace, thereby simplifying the analysis. We will adopt this construction to define suitable random matrix product operators in Section~\ref{sec:RMPOmodel}.} \\
\new{We now expand the modified overlap matrix $\tilde{G}$ as follows}
\begin{align}\label{eq:noise_1}
    \begin{split}
& \tilde{G}_{\sigma\pi}(q,\gamma) = \\
&= G_{\sigma \pi}(q) + \noisestrength \sum_{a=1}^k \llangle\sigma|\Id^{\otimes (a-1)} \otimes \tilde{\Noise}_a \otimes \Id^{\otimes (k-a)} |\pi\rrangle_q
+ O(\noisestrength^2) \, ,
    \end{split}
\end{align}
\new{where $\tilde{\Noise}$ (defined in Eq.~\eqref{eq:noise_tilde}) represents the first-order action of noise in $\noisestrength$, $\Id$ is the identity super-operator, and the index $a \in \{ 1, \dots, k \}$ labels the replicas.} \\

\new{As before, we consider a regime in which both $q^{-1}$ and $\noisestrength$ are small (and of the same order). In this regime, the term}
\begin{equation}\label{eq:ntilde_ele}
    \llangle \sigma | \Id^{\otimes (a-1)} \otimes \tilde{\Noise}_a \otimes \Id^{\otimes (k-a)} | \pi \rrangle_q ,
\end{equation}
\new{appearing in the second line of Eq.~\eqref{eq:noise_1}, contributes only to diagonal matrix elements, i.e.\ only when $\sigma = \pi$. Indeed, off-diagonal terms ($\sigma \neq \pi$) are suppressed by at least a factor $q^{-1}$ relative to the diagonal ones, since the super-operator $\tilde{\Noise}_a$ in Eq.~\eqref{eq:ntilde_ele}, as it acts only on one replica, cannot restore the reduced overlap between $\sigma$ and $\pi$ (recall that for any $\sigma \neq \pi$, $G_{\sigma \pi}\, q^{-k} = O(q^{-1})$). Therefore, for $\sigma\neq \pi$ the second term in Eq.~\eqref{eq:noise_1} is subleading, specifically $O(\gamma q^{-1})$.
As a result, we only need to evaluate the diagonal matrix elements $\llangle\sigma|\Id^{\otimes (a-1)} \otimes \tilde{\Noise}_a \otimes \Id^{\otimes (k-a)} |\sigma\rrangle_q$. To visualize these elements, we set $k = 4$, consider the specific permutation $\sigma = (1)(2\ 3\ 4)$, and use a graphical notation. If the noise channel $\tilde{\Noise}$ is placed on the first replica, i.e.\ the $a = 1$ term in the sum, then we have}
\begin{equation}\label{eq:noise_perm_1}
\raisebox{0.94cm}{$\llangle\sigma|\tilde{\Noise} \otimes \Id^{\otimes 3} |\sigma\rrangle_q = \, \, \, $} \includegraphics[width=0.5\linewidth]{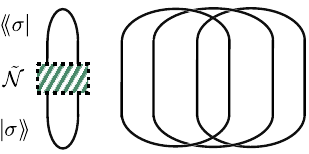} \raisebox{0.94cm}{$\, .$}
\end{equation}
\new{where the permutation $\sigma$ has been folded as in Eq.~\eqref{eq:perm_neww}. We observe that since $a=1$ corresponds to a fixed point of $\sigma$, i.e.\ $\sigma(a)=a$, Eq.~\eqref{eq:noise_perm_1} contains the factor}
\begin{equation}\label{eq:noise_fp}
     \includegraphics[width=0.18\linewidth]{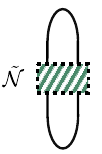} \raisebox{1.05cm}{$= \llangle \idp | \tilde{\Noise} |  \idp \rrangle_q = \Tr[\tilde{\Noise}(\Id)] = 0 \, .$}
\end{equation}
\new{Therefore, all fixed points of $\sigma$ give a vanishing contribution to the sum in Eq.~\eqref{eq:noise_1}. Conversely, when the noise channel $\tilde{\Noise}$ acts on a replica that is not a fixed point of $\sigma$, i.e.\ $\sigma(a)\neq a$, non-vanishing contributions arise. For example, setting $a=2$ we obtain the matrix element}
\begin{equation}\label{eq:noise_perm_2}
\raisebox{0.94cm}{$\llangle\sigma| \Id \otimes \tilde{\Noise} \otimes \Id^{\otimes 2} |\sigma\rrangle_q = \, \, \, $} \includegraphics[width=0.5\linewidth]{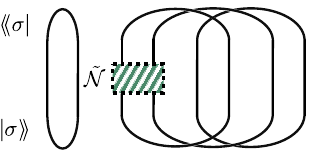} \raisebox{0.94cm}{$\, .$}
\end{equation}
\new{This term is proportional to}
\vspace{-0.5 cm}
\begin{align}\label{eq:KK}
    \begin{split}
      \includegraphics[width=0.18\linewidth]{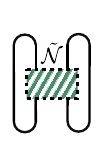} & \raisebox{1.05cm}{$= \sum_{\alpha \geq 1 } \Tr[\tilde{K}_\alpha] \Tr[\tilde{K}_\alpha^\dag] - q \, \Tr[\tilde{K}_0] - q \, \Tr[\tilde{K}_0^\dag] = $} \\[-2.3em]
      & \, \, =  -q \, \sum_{\alpha \geq 1 } \Tr[\tilde{K}_\alpha^\dag \tilde{K}_\alpha] \, ,
    \end{split}
\end{align}
\new{where we used that the $\tilde{K}_\alpha$ are traceless, together with the trace-preserving condition for $\tilde{K}_0$. The proportionality constant can be determined by noting that the noise channel acts as the identity on the remaining $k-2$ lines, yielding a factor $q^{k-2}$. For instance, Eq.~\eqref{eq:noise_perm_2} (with $k=4$) includes the term in Eq.~\eqref{eq:KK} along with two additional loops, each corresponding to a free index summed over $q$ values, giving a factor $q^2=q^{k-2}$. Combining these observations and defining}
\begin{equation}
   \kappa = q^{-1} \sum_{\alpha \geq 1 } \Tr[\tilde{K}_\alpha^\dag \tilde{K}_\alpha] \, ,
\end{equation}
\new{we obtain}
\begin{align}
\label{eq:sigmanoise}
    \begin{split}
    &\sum_{a=1}^k  \llangle\sigma|\Id^{\otimes (a-1)} \otimes \tilde{\Noise}_a \otimes \Id^{\otimes (k-a)} |\sigma\rrangle_q = \\
    &= - q^{k} \kappa \sum_{a=1}^k (1-\delta_{a,\sigma(a)})  \\
    &= - q^{k} \kappa (k - \mathrm{n_F}(\sigma)) \, .
    \end{split}
\end{align}
\new{Notice that $\kappa$ is necessarily positive, since $\tilde{K}_\alpha^\dag \tilde{K}_\alpha$ is positive. Moreover, using the freedom to rescale the $\tilde{K}_{\alpha}$ (see Section~\ref{sec:sub_noise}), one can always set $\kappa=1$, which we will do in the following.}
\new{Including the expansion of the first term in Eq.~\eqref{eq:noise_1}, namely
$G_{\sigma \pi}(q) = q^{k} \left( \delta_{\sigma \pi} + q^{-1} A_{\sigma \pi} + O(q^{-2}) \right)$ (see Eq.~\eqref{eq:expanding_GWG}), which provides corrections to the off-diagonal terms, we obtain}
\begin{align}\label{eq:G_tilde_final}
    \begin{split}
&  \tilde{G}(q, \noisestrength)= q^{k} \bigg(1 + \frac{A}{q} - \gamma \, \PP  + O(q^{-2}, \gamma^2, q^{-1} \gamma )\bigg) \, .
    \end{split}
\end{align}
\new{Finally, using Eq.~\eqref{eq:G_tilde_final} and Eq.~\eqref{eq:expanding_GWG}, we find}
\begin{align}\label{eq:expanding_noisy_W}
    \begin{split}
 &\tilde{\Wg}(q,\noisestrength) = \Wg(q) \, \tilde{G}(q,\noisestrength) \, \Wg(q) = \\
 &=q^{-k} \left( \Id - \frac{A}{q} - \noisestrength \, Q + O(q^{-2},\gamma^2,q^{-1} \noisestrength)\right) \, .
    \end{split}
\end{align}
\new{This form is identical to the one derived for depolarizing noise (Eq.~\eqref{eq:noise_wg_exp}). Working with generic Kraus channels thus confirms the universal validity of the weak-noise expansion for $\noisestrength \ll 1$ and large $q$ (with $q^{-1}$ and $\noisestrength$ taken to be of the same order).}


\subsection{Universality of the weak-noise expansion}
\new{While the derivation of the modified Weingarten coefficients assumed a single noise parameter $\gamma$, it extends straightforwardly to more general noise channels depending on an arbitrary number of parameters in the weak-noise limit, i.e.\ when all parameters are small enough and the channel remains close to the identity super-operator. As a concrete example, consider a noise channel characterized by two microscopic parameters $\noisestrength_1$ and $\noisestrength_2$, obtained as the successive action of two noise maps:
$\Noise_{\noisestrength_2, \noisestrength_1} \left(\rho \right) = \Noise_{\noisestrength_2}^{(2)} \left( \Noise_{\noisestrength_1}^{(1)} \left(\rho \right) \right)$.
Assuming that $\noisestrength_1$ and $\noisestrength_2$ are both small and of the same order, we can expand the channel to linear order as}
\begin{equation*}
\Noise_{\noisestrength_2, \noisestrength_1} \left(\rho \right) = \rho + \noisestrength_1 \tilde{\Noise}^{(1)}(\rho) + \noisestrength_2 \tilde{\Noise}^{(2)}(\rho) + O(\noisestrength_1^2, \noisestrength_1 \noisestrength_2, \noisestrength_2^2) \,.
\end{equation*}
\new{Repeating the same steps as before, one arrives at the modified overlap matrix
$\tilde{G}_{\sigma \pi}(q,\gamma) = \langle\!\langle \sigma| \Noise_{\noisestrength_1, \noisestrength_2}^{\otimes k} |\pi\rangle\!\rangle_q$, whose first-order correction is now}
\begin{equation}
    \sum_{a=1}^k \llangle \sigma | \Id^{\otimes (a-1)} \otimes \big( \gamma_1 \tilde{\Noise}_a^{(1)} + \gamma_2 \tilde{\Noise}_a^{(2)} \big)\otimes \Id^{\otimes (k-a)} | \pi \rrangle_q \, .
\end{equation}
\new{Since the two noise contributions combine linearly at this order, the previous derivation applies independently to the terms involving $\tilde{\Noise}_a^{(1)}$ and $\tilde{\Noise}_a^{(2)}$. This again leads to Eq.~\eqref{eq:G_tilde_final} and Eq.~\eqref{eq:expanding_noisy_W}, with an effective noise parameter given by $\noisestrength = \noisestrength_1 + \noisestrength_2$. We therefore conclude that our weak-noise treatment remains valid whenever each source of noise is associated with a small strength $\noisestrength_i$, a physically realistic scenario in several experimental settings.}
%


 \section{Random Matrix Product Operator Model and universality}
 \label{sec:RMPOmodel}

\new{We now consider the moments of the PoP distribution, i.e.\ the IPRs, at finite depth, $t\sim \log N$, in generic chaotic circuits. 
We begin by summarizing a general argument showing that, in the absence of noise, the calculation of these moments can be reduced to that of an equivalent one-dimensional problem. To this purpose, let us consider a random brick-wall circuit composed of independently Haar-distributed local gates. The computation of the $k$-th IPR involves $k$ replicas of the system, and performing the Haar average over each local gate (via repeated use of Eq.~\eqref{eq:weingarten}) replaces the original sum over microscopic degrees of freedom (qudits) with a sum over effective degrees of freedom taking values in permutations~\cite{Fisher_2023,PhysRevX.7.031016,Nahum2018Operator}. In this way one obtains a classical, Ising-like statistical mechanics model defined on a rectangular spacetime domain (of linear sizes $N$ and $t$), whose local variables are permutations and whose Boltzmann weights are given by Weingarten functions $\Wg$ and overlap $G$ matrices. The resulting interaction is generically \emph{ferromagnetic} (see, e.g., Eqs.~\eqref{eq:expanding_GWG}), and therefore favors the formation of spatiotemporal domains characterized by a uniform permutation (see e.g.~\cite{Nahum2018Operator,PhysRevB.101.104302,PhysRevX.7.031016,PhysRevB.99.174205,PhysRevX.10.031066,ChanDeLucaChalker2018}).} 

\new{For shallow circuits, where $t \ll N$, one can envision a coarse-graining procedure in which all microscopic degrees of freedom below a scale $\sim t$ are decimated. In the spirit of renormalization group theory, this procedure introduces new interactions among the remaining degrees of freedom. Following conventional arguments, we know that such renormalization flows push the system \emph{deeper into the ferromagnetic phase}. Consequently, in the coarse-grained picture, the model reduces to a one-dimensional strip consisting of distinct ferromagnetic domains, each characterized by a single permutation state (see Fig.~\ref{fig:1}(b)).} \new{The only remaining parameters are the \emph{interface cost} between adjacent domains carrying different permutations $\sigma, \, \sigma'$, which we denote $C_{\sigma \sigma'}(t)$. 
That such a description is possible ultimately follows from the \emph{membrane picture}~\cite{Fisher_2023,PhysRevB.99.174205,PhysRevX.10.031066,deluca2023universality}. 
As we will see explicitly in the next section, universality arises because, under very general assumptions~\cite{deluca2023universality}, the minimal-cost interfaces are the \emph{elementary} ones, namely those between permutations $\sigma$ and $\sigma'$ that differ by a single transposition (so that $A_{\sigma\sigma'}=1$). 
Defining $\Nth(t)\sim C_{\sigma\sigma'}^{-1}(t)$ for such pairs (the precise choice being immaterial), and taking the scaling limit $N,t\to\infty$ at fixed $x=N/\Nth(t)$, all other interfaces become negligible. 
One is then left with a dilute gas of elementary interfaces with fugacity $\Nth(t)^{-1}$. This, therefore, also clarifies why the half-chain purity, corresponding to the cost of the interface between the swap $\swp=(12)$ and the identity $\idp$ permutations, provides a direct measure of $\Nth(t)^{-1}$~\cite{Shivam_2023,Lami_2025,Christopoulos_2025}.  }


\new{Building on this perspective, in the next section we will evaluate these moments in a simpler class of quantum circuits, namely \emph{staircase circuits} (see, e.g., Eq.~\eqref{eq:mps_circuit_2}). Despite their simplicity, staircase circuits retain the same coarse-grained features and therefore allow for a direct calculation of universal quantities, while also enabling a controlled inclusion of noise. Without noise, these circuits are known to correspond to the sequential generation of one-dimensional tensor networks, namely Matrix Product States (MPS)~\cite{Schon_2007,garnerone2010typicality,PRXQuantum.2.040308,PRXQuantum.6.010345,Collura_2024}. Their bond dimension (i.e.\ the matrix size) $\chi$ is uniform (except near the boundaries) and is set by the support of the unitary gates (for instance, a staircase circuit featuring gates acting on $r+1$ qubits generates an MPS of bond dimension $\chi=2^r$). This construction can be straightforwardly extended to mixed states, where it generates Matrix Product Operators (MPOs), as we will see later. By introducing randomness, for instance in the form of Haar distributed unitary gates, one can use this constructions to naturally define notions of Random Matrix Product States (RMPS) and Random Matrix Product Operators (RMPOs)~\cite{garnerone2010typicality,PRXQuantum.2.040308,PRXQuantum.6.010345}. 
While we use the staircase architecture for explicit calculations, our conclusions extend beyond it: we will show that the resulting behavior is in fact \emph{universal}. To corroborate this claim, in Appendix~\ref{sec:RPM}, we analyze a different solvable random-circuit model, specifically the \textit{Random Phase Model} (RPM), and demonstrate that it exhibits the same universal properties.}

\subsection{Universal distribution in noiseless circuits}

We now summarize the derivation of the universal form of the moments of the PoP for generic noiseless chaotic circuits. One way to understand this is to represent the time-evolved state $U|\pmb{0}\rangle$ as an equivalent random matrix product state (RMPS). Following the discussion of the previous section, the bond dimension $\chi$ must be chosen in order to reproduce the exponential decay of the purity. This implies a bond dimension $\chi$ growing exponentially with the circuit depth $t$. More precisely, the Thouless length defined in Eq.~\eqref{eq:def_x} can be identified with the MPS bond dimension, $\Nth(t) \sim \chi d/(d-1)$
leading to the scaling limit~\cite{Lami_2025}
\begin{equation}\label{eq:x_def}
    x = \frac{d-1}{d} \frac{N}{\chi} \, .
\end{equation}
Then, the $k$-replica average expresses the IPRs as the partition function of a one-dimensional statistical-mechanical model with local degrees of freedom in permutation space~\cite{Christopoulos_2025, Lami_2025, Sauliere_2025}. Such a partition function can be expressed as a product of transfer matrices of size $k! \times k!$, ultimately related, in the scaling limit, to the matrix $A$ introduced above, leading to~\cite{Lami_2025}
\begin{equation}\label{eq:no_noise}
I_k \scaleq D^{1-k} 
(\pmb{1} | e^{x A} | \pmb{1} )
= I_k^{\text{PT}} \,  \exp(x \frac{k(k-1)}{2}) \, .
\end{equation}
\new{Here, we use a bra $(\dots|$ and ket $|\dots)$ notation to indicate vectors of length $k!$ with components associated to permutations and $|\pmb{1})=(1, 1, ...,1)$ denotes the vector with all unit components of length $k!$.} \new{This is also the eigenvector of the matrix $A$ with maximal eigenvalue, namely $k(k-1)/2$, which explains the second equality in Eq.~\eqref{eq:no_noise}.
This expression confirms in quantitative terms the discussion in the previous section: 
the exponential factor is the partition function of a gas of non-interacting 
domain walls, corresponding to the transpositions. 
The parameter $x = N/\Nth(t)$ emerges from the balance between fugacity and system size, 
and plays the role of the Poisson intensity controlling the number of domain walls, 
thereby leading to the characteristic exponential form of the partition function. 
As a paradigmatic example, the case $k=2$ has a complete analogy with the classical 1D Ising model, since $S_2 \simeq \mathbb{Z}_2$ as it is evident by mapping the two permutations identity $\idp$ and swap $\swp$ to standard Ising variables $\pm 1$. In this case, domain walls have a Boltzmann cost $e^{-\beta J} = \Nth(t)^{-1} \sim e^{-t/\tau}$, with $\beta$ the inverse temperature and $J$ the Ising coupling. Therefore, in the scaling limit Eq.~\eqref{eq:def_x}, where $t \sim \log N \to \infty$, this corresponds to very low temperature, in which domain walls are \textit{diluted} but their cost is balanced by the entropic factor $N$.
}
\\
Finally, one can reconstruct the PoP distribution at any $x$ by recognizing that the moments in Eq.~\eqref{eq:no_noise} correspond to those of a product of two independent random variables, namely $w = w_{\rm PT} \, g$, where $w_{\rm PT}$ is drawn from the Porter--Thomas distribution and $g$ from a log-normal one (such that $\mathbb{E}[g^k] = \exp(x \, k(k-1)/2)$). As an explicit confirmation of the emergent universality, these results coincide perfectly with those obtained for the RPM in the limit of large local Hilbert-space dimension $d$~\cite{Christopoulos_2025}.

\subsection{Universal distribution in noisy circuits}

The ideas and results presented above can be successfully extended to circuits in the presence of noise, yielding quantitative predictions for finite-depth circuits. To facilitate the presentation of the argument, and following the previous argument regarding the reduction to effective one-dimensional models, we model a generic noisy circuit as a noisy staircase quantum circuit. Specifically, our construction is as follows:
\begin{equation}\label{eq:mps_circuit_2}
\includegraphics[width=0.35\linewidth, valign=c]{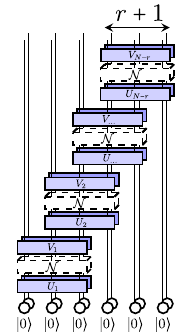} \, .
\end{equation}

\new{Each step of the staircase consists of a unitary gate $U_i$, followed by a noise channel $\mathcal{N}_{\noisestrength}$, and then a second unitary gate $V_i$, similarly to the global gates we discussed in Sec.~\ref{subsec:Generic noise}. Since $\mathcal{N}_{\noisestrength}$ acts on a density matrix, both the state and its conjugate appear in Eq.~\eqref{eq:mps_circuit_2}, with conjugate operators represented by darker colors. For now, we do not need to specify the precise form of $\mathcal{N}_{\noisestrength}$ beyond Eq.~\eqref{eq:noisechandef}. The gates $U_i$, $V_i$, and the noise channel act on $r+1$ qudits, each of local dimension $d$ (for illustrative purposes, $r=2$ in Eq.~\eqref{eq:mps_circuit_2}). The full set of gates is denoted by $U := \{U_i, V_i\}_{i=1}^{N-r}$. In the absence of noise, Eq.~\eqref{eq:mps_circuit_2} reduces to the standard sequential generation of an MPS with bond dimension $\chi$, starting from qudits initialized in the state $|0\rangle$~\cite{Schon_2007}. Specifically, the parameter $r$ is related to the MPS bond dimension $\chi$ through, as specifed also in the previous section, $ r = \log_d \chi$.}

The circuit of Eq.~\eqref{eq:mps_circuit_2} generates a mixed density matrix $\rho_{\Noise}(U)$, which can be represented as a \textit{Random Matrix Product Operator} (RMPO), see also Fig.~\ref{fig:1}. This becomes evident by suitably reshaping Eq.~\eqref{eq:mps_circuit_2} and \new{by fusing the gates $U_i$ and $V_i$ (as well as their conjugates) with the corresponding noise channels}. This yields
\begin{equation}
\includegraphics[width=0.45\linewidth, valign=c]{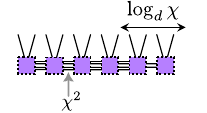} \, ,
\end{equation}
which is an MPO of bond dimension $\chi^2$, shown in its vectorized form (the bond dimension decreases by a factor $d^2$ after each of the last $r$ sites near the right boundary). \\
\noindent
To compute the overlap, Eq.~\eqref{eq:overlap0}, $w_{\boldsymbol{x}}=D\langle \boldsymbol{x}|\rho_{\Noise}(U)|\boldsymbol{x}\rangle=D \llangle \boldsymbol{x}, \boldsymbol{x}|\rho_{\Noise}(U) \rrangle$, the open physical legs in Eq.~\eqref{eq:mps_circuit_2} should be contracted with a bit string $|\boldsymbol{x}\rangle = |x_1, \dots, x_N \rangle$.
Afterwards, to evaluate $I_k = D^{-k} \sum_{\boldsymbol{x}} \mathbb{E}_U[w_{\boldsymbol{x}}^k]$, one should replicate this network $k$ times. \new{At this stage, the fundamental repeated block that needs to be evaluated can be represented as follows:
\begin{equation}
\includegraphics[width=0.45\linewidth, valign=c]{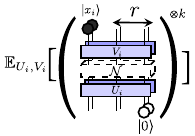} \, ,
\end{equation}
where the grey dots represent computational basis states $|x_i\rangle$.}
\new{Next, Eq.~\eqref{eq:noisy_wg_new_1} must be applied to compute the average $\mathbb{E}_{U_i, V_i \sim \Haar(d \chi)}[\dots]$. This enables us to express the IPR $I_k$ in terms of powers of an appropriate transfer matrix in replica space, i.e.\ in the $k!$-dimensional space spanned by the permutation states $|\sigma\rrangle$.}
Specifically, we arrive at (see details in Appendix~\ref{sec:app_singlequdit})
\begin{equation}\label{eq:contraction}
    I_k = D (L| T^{N-r-1} |R) \, ,
\end{equation}
where the transfer matrix
\begin{equation}\label{eq:TM}
    T := \tilde{\Wg}(d \chi,\noisestrength) \, G(\chi) \, ,
\end{equation}
acts in the space of permutations, and therefore has size $k! \times k!$.  The boundary vectors in Eq.~\eqref{eq:contraction} read $(L|=(\pmb{1}|$ and $|R) = \tilde{\Wg}(d\chi,\noisestrength) |\pmb{1})$. For small replica numbers $k$, Eq.~\eqref{eq:contraction} enables the exact evaluation of $I_k$ for arbitrary noise channels, via direct multiplication of matrices of size $k! \times k!$. \\

\new{We now focus on the regime in which the noise strength $\noisestrength$ decreases as the system size grows, according to a specific \emph{scaling limit}. The scaling limit (s.l.) is defined as follows: first, in accordance with the noiseless case discussed above and Eq.~\eqref{eq:x_def}, we keep $x = N/\chi \cdot (d-1)/d$ fixed as $N, \chi \to \infty$; second, we scale the noise strength as}
\begin{equation}\label{eq:epsilon_def}
    \gamma = \frac{\eta}{N} \, .
\end{equation}
\new{In this regime, $\chi^{-1}$ and $\gamma$ are both $O(N^{-1})$, and we can therefore use the expansion of the noisy Weingarten coefficients derived in the previous section. Specifically, using Eq.~\eqref{eq:expanding_noisy_W} and Eq.~\eqref{eq:expanding_GWG}, we arrive at}
\begin{align}
    \begin{split}
    & T =  \\
    &= d^{-k} \left( \Id - \frac{x}{N} \frac{1}{d-1} A - \frac{\eta}{N} \, Q \right) \left( \Id + \frac{x}{N} \frac{d}{d-1} A \right) + O(N^{-2})= \\
    &= d^{-k} \left( \Id + \frac{x}{N} A - \frac{\eta}{N} \, Q \right) + O(N^{-2}) \, .
    \end{split}
\end{align}
\new{By using this equality and neglecting all terms of order $O(N^{-2})$, which are irrelevant in the scaling limit, we finally obtain the following expression for the transfer matrix}
\begin{equation}\label{eq:T_final}
T \scaleq d^{-k} \exp(\frac{x}{N} A - \frac{\eta}{N} \, \PP) \, .
\end{equation}
We stress that our weak-noise expansion leading to Eq.~\eqref{eq:T_final} is completely \emph{independent of the specific details of the noise, including whether the channel $\mathcal{N}$ is unital or not}. Using this equation and taking the scaling limit, Eq.~\eqref{eq:contraction} reduces to
\begin{equation}\label{eq:TM_sl}
    I_k \scaleq 
    I_k(x,\eta) := D^{1-k} 
    (\pmb{1} | e^{x A -\eta \PP} | \pmb{1} ) \
    \, .
\end{equation}
\new{
The form~\eqref{eq:TM_sl} for the IPRs is one of the main results of this paper. 
Consistently, it reduces to the noiseless case, Eq.~\eqref{eq:no_noise}, when $\eta=0$. 
As discussed above, the resulting expression admits a natural interpretation in terms of 
the partition function of a one-dimensional gas of domain walls in permutation space. 
In this picture, the inverse Thouless length $L(t)^{-1}$ plays the role of a fugacity, while the parameter $x=N/\Nth(t)$ corresponds to the associated extensive Poisson intensity controlling the total number of domain walls. The noise parameter $\eta$ introduces an additional bias, effectively acting as a magnetic field that favors permutations with a larger number of fixed points, i.e.\ configurations 
closer to the identity $\idp$. For $k=2$, the analogy with the classical one-dimensional Ising model is once again useful. In this case, the matrix $A$ reduces to the Pauli matrix $\PauliX$, while $Q = \mathrm{diag}(0,2)$, making explicit its connection to the local magnetization and the interpretation of $\eta$ as a magnetic field.}

The universality of Eq.~\eqref{eq:TM_sl} rests on two facts: first, as in the noiseless case, in the scaling regime, 
the domain walls are effectively diluted and only the elementary ones (transpositions) associated with the matrix $A$ are relevant; second, while noise generally favors the identity permutation, the generic measure of distance from the identity emerging in the weak-noise expansion is simply the number of fixed points. \new{The latter is a consequence of the fact that any effect of weak, local noise manifests itself as the insertion of a quantum channel close to identity; in the analysis of replicas, it manifests itself as a term immersed in a ferromagnetic domain of a certain permutation $\sigma$; we can therefore use the expression Eq.~\eqref{eq:sigmanoise} which, in the scaling limit, leads to \eqref{eq:TM_sl}.
From these considerations, it also follows that $\eta$ is a bulk property that does not change the cost $x$ associated with each interface, which therefore remains the same as in the model without noise.}
Thus, only two parameters $x$ and $\eta$ encode all microscopic interactions and the specific form of the noise, where $x$ is still expressed in terms of the Thouless length (Eq.~\eqref{eq:x_def}), in turn linked to the half-chain purity of the noiseless model. This hypothesis is verified in the models we solve analytically and in the numerical data we present below. Note that in our RMPO model, Eq.~\eqref{eq:mps_circuit_2}, approximately $r$ noise channels are applied per site. As a result, the total number of noisy gates scales as $\Ng \propto N$, and $\noisestrength = \eta / N$. In a realistic quantum circuit extended in time, the number of errors is naturally proportional to the circuit depth. In that case, the time variable $t$ plays the role of $r$, leading to $\Ng \sim N t$. The relevance of our results for this case will be addressed in Section~\ref{sec:strong_noise}.

The PoP $P_{x,\eta}(w)$ is uniquely determined by the moments in Eq.~\eqref{eq:TM_sl}. In the noiseless case ($\eta=0$), Eq.~\eqref{eq:TM_sl} reduces to Eq.~\eqref{eq:no_noise} and $P_{x,0}(w)$ is a log-normal convolved with PT. Moreover, for infinite circuit depth ($x = 0$), one recovers the moments of the SPT distribution in Eq.~\eqref{eq:shiftedPT} (see the previous section). Thus, Eq.~\eqref{eq:TM_sl} interpolates between these two regimes, capturing the combined effects of noise and finite depth through the scaling parameters $x$ and $\eta$. For small $k$, Eq.~\eqref{eq:TM_sl} provides an efficient way to compute the scaling form of the IPRs for arbitrary $x$ and $\eta$. In particular, in the same framework and scaling limit, we can derive a closed form for $\xeb$, as it is closely related to $I_2$. Technically, the discrepancy between $\xeb$ and $I_2$ lies in the fact that in Eq.~\eqref{eq:xebdef} only one of the two replicas contains noise, thus leading to
\begin{align}\label{eq:xeb_sl}
\begin{split}
      \xeb &\scaleq (\pmb{1}| e^{x A - \frac{\eta}{2} \PP} |\pmb{1}) - 1 = \\ &=
      2 e^{-\eta/2} \left[\cosh \left( \theta(x,\eta) \right)
       + \frac{x\,\sinh \left( \theta(x,\eta) \right)}
              {\theta(x,\eta)}\right] - 1 \, ,
\end{split}
\end{align}
where $\theta(x,\eta)=\sqrt{x^2 + (\eta/2)^2}$ \new{and $\eta$ appears divided by a factor of two compared to Eq.~\eqref{eq:TM_sl}, since noise affects only one of the two replicas}. \new{The second line of Eq.~\eqref{eq:xeb_sl} follows directly from the first by diagonalizing the $2\times 2$ symmetric matrix $x A- \frac{\eta}{2} \PP$ and computing the overlap of the eigenvectors with the boundary vector $|\pmb{1})=(1,1)$.} For $x=0$, $\xeb$ reduces to the circuit fidelity $F$, as expected from the white-noise approximation that becomes valid in the weak-noise limit~\cite{Dalzell2024}. In fact, we find (see Appendix~\ref{sec:app_singlequdit} for the explicit calculation of the average fidelity):
\begin{equation}\label{eq:xeb_at_zero}
    \lim_{x \to 0^+} \xeb = e^{-\eta}=F  \, .
\end{equation}
This relation shows that, in general, the parameter $\eta$ is fixed by the global fidelity of the circuit. The parameter $x$ is fixed by the system size and by the spreading of purity in the noiseless circuit (see Eq.~\eqref{eq:def_x} and Fig.~\ref{fig:Purity}). In the case of a one-dimensional Haar-random brickwall circuit with $d$-dimensional qudits, the scaling of the \new{half-chain} purity is explicitly known~\cite{PhysRevX.7.031016}, and is given by
\begin{equation}\label{eq:tau}
    \tau_{\text{bw}}^{-1}(d) = \log \left(   \frac{d^2 + 1}{2d} \right) \, ,
\end{equation}
which for qubits gives $\tau_{\text{bw}}^{-1}(2) \simeq 0.223$.

In contrast to the $\eta=0$ case, an analytic evaluation of the moments in Eq.~\eqref{eq:TM_sl} for arbitrary $k$---and thus an explicit reconstruction of the distribution---is technically challenging. Nonetheless, the behavior of the IPRs in different regimes of $x$ can be extracted perturbatively, as we show below. Additionally, we provide an efficient numerical procedure to extract the PoP.

\begin{figure*}[t!]
    \centering
    \includegraphics[width=1.\linewidth]{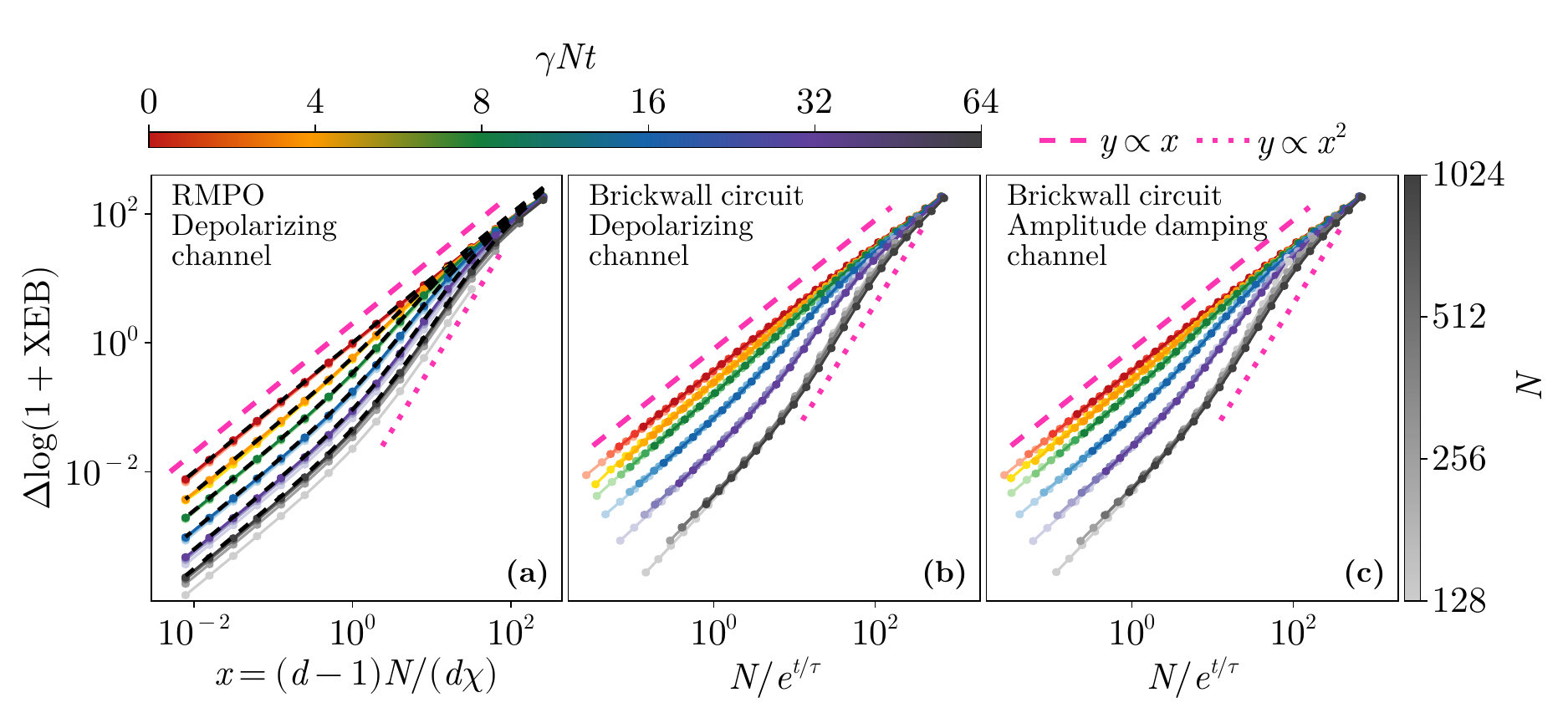}
    \caption{Plot of the $\xeb$ for noisy RMPO (a), and random brickwall circuits (b)--(c). Different noise models are considered: RMPO with depolarizing channel (a), see Eq.~\eqref{eq:mps_circuit_2}, and random brickwall circuits with two-qubit depolarizing noise (b), or single-qubit amplitude-damping noise (c). The noise strength scales as $\noisestrength  \propto 1/(N r)$ for RMPO (a), and as $\noisestrength  \propto 1/(N t)$ for the circuits (b)--(c). The proportionality constant is indicated by the horizontal color bar at the top. Note that in panel (a) the label is $\noisestrength N r$ rather than $\noisestrength N t$. We plot $\Delta\log(1+\xeb)=\log(1+\xeb)-\lim_{x \to 0^+}\log(1+\xeb)$ as a function of the rescaled circuit depth $x \sim N e^{- t/ \tau}$. We identify three different scaling regimes fixed by the fitting parameter $\eta = - \log F$. For $x \gg \eta$, the scaling is with $x$ and is independent of $\eta$, showing that the system is effectively noiseless. For $x^2\sim \eta$, the scaling is with $x^2$, characteristic of strong noise. For $x<1$, it approaches its asymptotic value linearly in $x$. In panel (a) we also plot the prediction of Eq.~\eqref{eq:xeb_sl} (dashed black line).
    }
    \label{fig:XEBplot}
\end{figure*}
\section{The three dynamical regimes}
For fixed $\eta$, Eqs.~\eqref{eq:TM_sl} and \eqref{eq:xeb_sl} reveal three distinct dynamical regimes as the circuit depth, and hence $x$, is varied, and which are manifested from the scaling of the $\xeb$, see Fig. \ref{fig:XEBplot}. These can be naturally grouped into a short-depth case and an intermediate-to-long-depth one. Notice that, for practical reasons, we define the deviation from the asymptotic value of the $\xeb$ in log-scale as $\Delta \log(1+\xeb):= \log(1+\xeb) - \lim_{x \to 0^+}\log(1+\xeb)$.  \\

    \textit{Short‐depth $x \gg \eta$}. A simple inspection of Eq.~\eqref{eq:xeb_sl} shows that $\Delta \log(1+\xeb) \propto x$ in this regime. Instead, Eq.~\eqref{eq:TM_sl} can be analyzed using perturbation theory in the small parameter $\eta/x$, expanding around the noiseless limit $\eta = 0$ (Eq.~\eqref{eq:no_noise}). At first order, we find $I_k \simeq I_k^{\text{PT}} \exp(x \frac{k(k-1)}{2} - \eta(k-1) + O(\eta^2/x))$. The fact that the right-hand side of this expression does not yield $\lim_{k \to 0} I_k = D$, as it trivially should according to the definition in Eq.~\eqref{eq:definitions}, indicates that the first-order expansion at small $\eta/x$ cannot be analytically continued to $k < 1$. Nevertheless, it allows us to extract the PoP tails at $w\gg 1$ as a convolution of the PT with a (unnormalized) lognormal. \\

    \textit{Large/intermediate‐depth}. To simplify the analysis, we assume that $\eta$ is large enough that we can focus in practice only on the ground state of $Q$, which consists of the identity permutation $e$, perturbed by the matrix $A$. Without perturbation, and neglecting terms of order $O(e^{-2\eta})$, one simply obtains the classical value of the IPR, $I_k^{\text{C}}$. The perturbation introduces two relevant types of corrections. First, from second-order perturbation theory, one can obtain the correction: $x^2/(2\eta)\sum_{\sigma} A_{e,\sigma}^2 = x^2 k(k-1)/(4 \eta)$. This expression can be interpreted as describing a region in the bulk where the state is locally excited to $\sigma$ instead of remaining in the ground state $e$. This excited region is bounded by two domain walls, corresponding to identical transpositions that separate it from the surrounding ground state regions. Pictorially, we can represent this correction as follows:
\begin{equation}
\begin{tikzpicture}[baseline=(current  bounding  box.center),scale=0.8]
\definecolor{mycolor1}{rgb}{0.0, 0.83, 0.39}
\definecolor{mycolor2}{rgb}{0.83, 0.0, 0.39}
\pgfmathsetmacro{\ll}{1.25}
\draw[line width=0.8mm, mycolor1, ultra thick] (-2*\ll,0) -- (-\ll/2,0.);
\draw[line width=0.8mm, mycolor1, ultra thick] (2*\ll,0) -- (\ll/2,0.);
\draw[line width=1.mm, mycolor2] (-\ll/2,0.) -- (+\ll/2,0.);
\node[scale=1.] at (0,-0.35) {$\sigma$};
\node[scale=1.] at (-5/4*\ll,-0.35) {$e$};
\node[scale=1.] at (+5/4*\ll,-0.35) {$e$};
\end{tikzpicture}
\end{equation}
     Second, the boundary state $|1)$ in Eq.~\eqref{eq:TM_sl} contains contributions from all permutations. As a result, transpositions $\sigma$ can remain localized near the left or right boundaries before the perturbation given by the $A$ matrix causes them to `jump' to the ground state $e$, creating a domain wall at cost $O(x)$. This correction can be pictorially represented as
\begin{equation}
\begin{tikzpicture}[baseline=(current  bounding  box.center),scale=0.8]
\definecolor{mycolor1}{rgb}{0.0, 0.83, 0.39}
\definecolor{mycolor2}{rgb}{0.83, 0.0, 0.39}
\pgfmathsetmacro{\ll}{1.25}
\draw[line width=0.8mm, mycolor1, ultra thick] (-2*\ll,0) -- (\ll,0.);
\draw[line width=1.mm, mycolor2] (2*\ll,0) -- (\ll,0.);
\node[scale=1.] at (-\ll/2,-0.35) {$e$};
\node[scale=1.] at (3/2*\ll,-0.35) {$\sigma$};
\end{tikzpicture}
\end{equation}   
{and similarly on the left boundary}. Putting these two contributions together, we arrive at
\begin{equation}
    I_k = I_k^{\text{C}} \exp(\frac{x^2}{\eta}\frac{k(k-1)}{4}) \Big(1+\frac{x}{\eta} \frac{k(k-1)}{2} + O(x^2/\eta^2) \Big) .
\end{equation}
The first term of this expression again indicates a log-normal PoP (the convolution with the delta distribution $P_{\text{C}}(w)$ is irrelevant). This expression also determines two distinct regimes by increasing the depth of the circuit (decreasing $x$). When $1 \ll x = \sqrt{\eta} \ll \eta$, the exponential factor dominates giving rise to $\log \mathbb{E}[w^k] \propto x^2/\eta$ behavior. In contrast, at larger depths for which $x\ll 1$, $\log \mathbb{E}[w^k] \propto x/\eta$. This behavior is explicitly confirmed by the analytical expression of the $\xeb$ \eqref{eq:xeb_sl}. \\

The existence of these dynamic regimes is a direct consequence of the universal form \eqref{eq:TM_sl}. Fig.~\ref{fig:XEBplot} presents numerical results for both noisy RMPO and random brickwall circuits confirming this analysis. For the brickwall circuit, we consider two types of noise: a two-qubit depolarizing channel and a single-qubit amplitude damping channel
$\mathcal{N}_{\noisestrength}^{\text{amp.}}(\rho) = K_0\rho K_0^\dagger +  K_1\rho K_1^\dagger$ with $K_0=\begin{pmatrix}
    1 & 0\\0 & \sqrt{1-\noisestrength}
\end{pmatrix}$ and $K_1=\begin{pmatrix}
    0 & \sqrt{\noisestrength}\\0 & 0
\end{pmatrix}$. 
Notably, the latter is non-unital, confirming that the emergence of these regimes does not depend on the unitality of the noise. We compute $\xeb$ by contracting the tensor network obtained by replicating the system and averaging all the unitary gates. The resulting replica tensor network lives in a $k!$-dimensional space, spanned by the permutations; for $k=2$, which is the case for $\xeb$, a basis is given by the identity $\idp$ and the transposition $\swp = (12)$. While for RMPS (and RMPO) the exact form of the rescaled depth $x= (N/\chi) (d-1)/d$ is derived analytically above, for the circuit we take the definition \eqref{eq:def_x} and fit the parameter $\Nth_0$ (the value of $\tau$ is provided in Eq.~\eqref{eq:tau} but it could also be fitted). We plot $\Delta \log(1+\xeb)$ as a function of $x$, or $N / e^{t / \tau}$ (circuit). For the circuit, we observe that plotting as a function of this variable yields good collapse of the curves corresponding to different system sizes $N$. Consistently, in all three cases, we clearly observe the predicted three regimes $O(x)$, $O(x^2)$ and $O(x)$. \\  

\begin{figure}[ht!]
    \centering \includegraphics[width=1\linewidth]{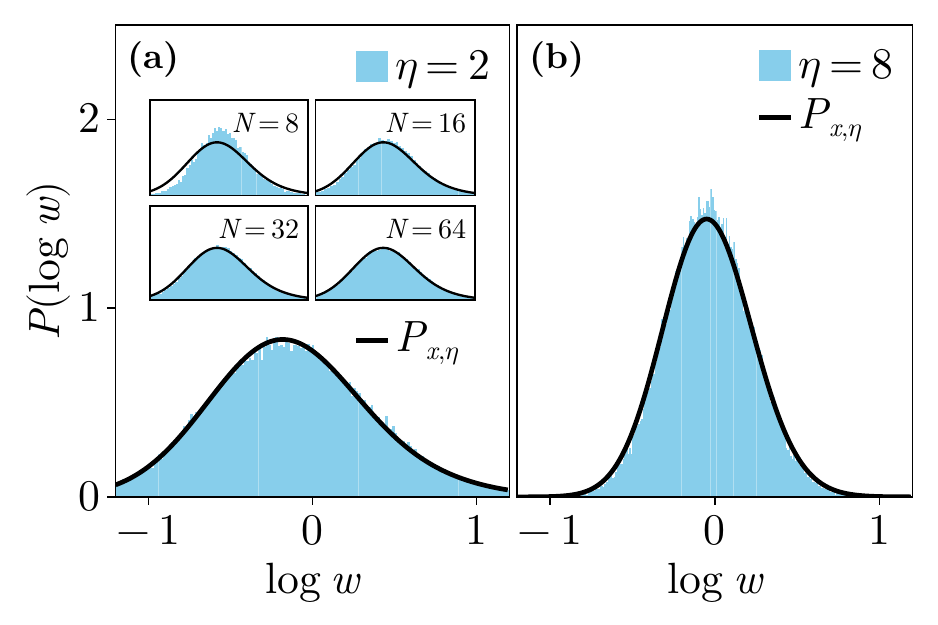}
    \caption{PoP of RMPO for construction of Eq.~\eqref{eq:mps_circuit_2} for $N=128$, $\chi=128$ ($x=0.5$), and $2\leq \eta \leq 8$. \new{For each set of parameters, the histogram is built from 100 independent circuit realizations, yielding a total of $10^5$ samples per histogram.} The distribution of $\log w$ can be approximated as a Edgeworth series (black line) where only the first two terms of the expansion have been considered and the cumulants of the distribution have been fitted from $I_k$ computed from Eq.~\eqref{eq:TM_sl} for $k\in\{1,...,6\}$. Inset of panel (a) shows the approach to the universal prediction by increasing $N$.}   \label{fig:rmpsdistribution}
 \end{figure}

\begin{figure*}[t!]
    \centering
    \includegraphics[width=\linewidth]{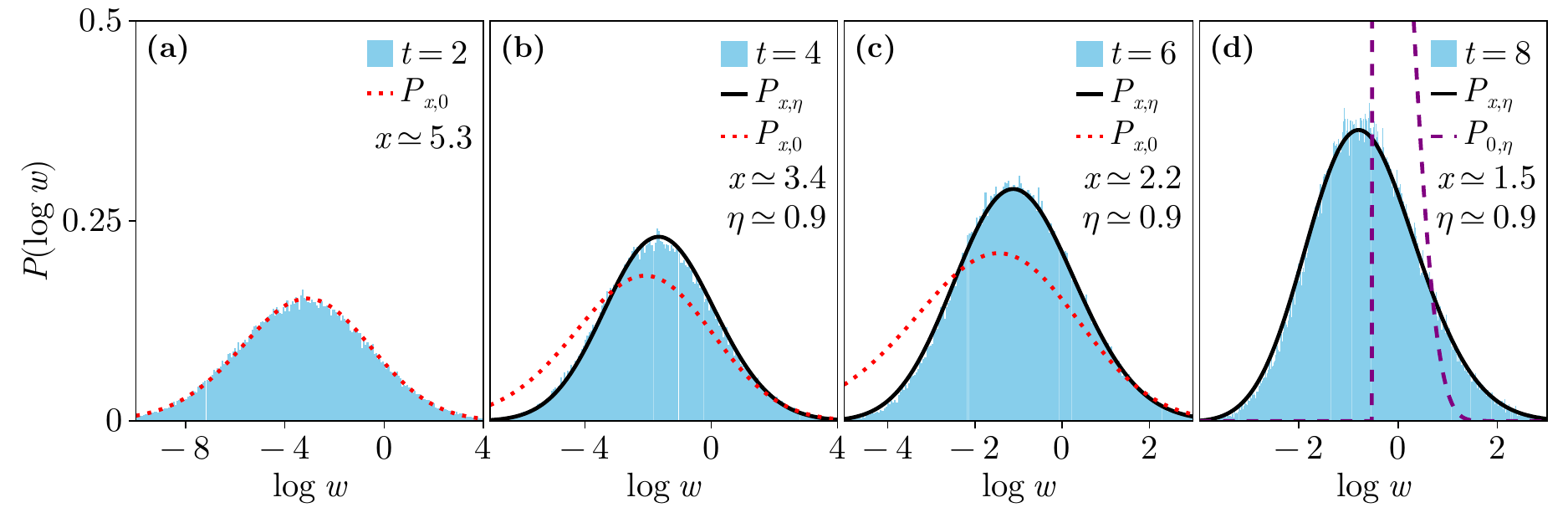}
    \caption{
    Plot of the PoP for a Haar-random brickwall circuit subject to two-qubit depolarizing noise with $\noisestrength N t = 2$ and system size $N = 32$. Different depths $t$ are explored (see values reported in the plots). \new{For each set of parameters, the histogram is built from 100 independent circuit realizations, yielding a total of $10^5$ samples per histogram.} At the shortest depth, the distribution is well approximated by the noiseless distribution (dotted red curve) $P_{x,0} \simeq P_{x,\eta}$, whereas the latter fails to accurately describe the data at larger depths. In contrast, the theoretically predicted distribution $P_{x,\eta}$ (black line) accurately reproduces the data. The shifted Porter-Thomas (SPT), Eq.~\eqref{eq:shiftedPT}, corresponding to the distribution at infinite times, $x=0$, is also plotted for reference in the last plot. The values of $x$ and $\eta$ are obtained by the fitting procedure involving the $\xeb$ data explained in the main text (Sec.~\ref{sec:pop}). }
    \label{fig:circuitdistribution}
\end{figure*}

\section{Universal form of the PoP distribution}\label{sec:pop}

While the derived form of the PoP $P_{x,\eta}(w)$ is fully determined by its moments in Eq.~\eqref{eq:TM_sl}, a fully analytical expression in closed form unfortunately remains inaccessible for generic $x$ and $\eta$. Nevertheless, using that for sufficiently large $\eta \gg x$, the distribution approaches a log-normal, namely a Gaussian in the variable $\log w$, allows for a systematic expansion. Specifically, we employ the Edgeworth series~\cite{hall1992bootstrap,Edgeworth1887TheLO}, which approximate a probability distribution in terms of its cumulants. \new{The latter can be computed numerically from Eq.~\eqref{eq:TM_sl} for sufficiently small $k$ (e.g., $k \in \{1,2,3,4,5,6\}$), since the sizes of the matrices $A$ and $\PP$ grow as $k!$.}
\new{To benchmark our results, in Fig.~\ref{fig:rmpsdistribution} we compare the distribution predicted by our formula using the Edgeworth series (black lines), with the empirical distribution of overlaps $w$ between random bit-strings and the RMPO, constructed as in Eq.~\eqref{eq:mps_circuit_2} (blue histogram). For this construction, we use the multi-qubit depolarizing channel.
We consider a specific value of $x$ and various values of $\eta$. We observe an excellent agreement between two distributions. Furthermore, in Fig.~\ref{fig:rmpsdistribution}(a), we illustrate how the data approach the Edgeworth expansion as the system size $N$ increases.} \\

\new{Far from being limited to this case, the PoP $P_{x,\eta}(w)$ is \emph{universal}}: it applies to generic circuit architectures and noise models, and is fully determined by the parameters $x$ and $\eta$. In generic cases, these should be regarded as model-dependent fitting parameters (while for RMPO their explicit form in terms of $\chi$, $N$, $d$ and $\noisestrength$ is known; see Eqs.~\eqref{eq:x_def} and \eqref{eq:epsilon_def}). \new{To show the universality of our form, in Fig.~\ref{fig:circuitdistribution} we compare our prediction Eq.~\eqref{eq:TM_sl} (black lines) with the empirical PoP distribution (blue histogram) for a \emph{Haar random brickwall circuit} with two-qubit depolarizing noise on each gate. The empirical overlaps} are obtained with tensor network simulations~\cite{vidal2004one,zwolak2004mixed,verstraete2004matrix,itensor}. We explore different values of depth $t$, while we fix $N$ and $\noisestrength N t$.
Extracting the exact values of $x$ and $\eta$ is nontrivial, as no analytical expressions are available for the brickwall circuit case.  
To determine these parameters, we therefore propose an experimentally friendly approach that can be applied to generic circuits at finite depth. The approach proceeds as follows. First, one computes $\xeb$ (which can also be measured experimentally) and extracts $\eta$ by extrapolating its value at large depth (ideally $x=0$, see Eq.~\eqref{eq:xeb_at_zero}). Next, $x$ can be determined by equating Eq.~\eqref{eq:xeb_sl} to the numerically or experimentally obtained $\xeb$ data. These parameters are then used to reconstruct the full distribution $P_{x,\eta}(w)$ with moments given by Eq.~\eqref{eq:TM_sl}. The fact that the values of $x$ and $\eta$ extracted solely from $\xeb$ yield good agreement is a nontrivial indication that all moments of the anticoncentration follow the same universal form, depending only on these two parameters. \new{In Appendix~\ref{ref:additionalnumerical}, this is demonstrated by reconstructing the $I_3$ moment from $\xeb$ (see Fig.~\ref{fig:I3}).}

In Fig.~\ref{fig:circuitdistribution}, the empirical distribution and our analytical prediction shows excellent agreement, with small deviations only due the finite size $N$ of the simulation. Dotted red line represents the noiseless distribution $P_{x,0}(w)$ (log-normal convolved with PT), which closely match the empirical distribution for short time $t$ (i.e.\ $x \gg 1$). At late time instead the distribution tends toward a shifted Porter Thomas, i.e.\ $P_{0,\eta}(w)$ (purple dashed line), \new{yet, Fig.~\ref{fig:circuitdistribution} shows that at intermediate times the two distributions remain clearly distinct, indicating that the shifted Porter–Thomas form should be regarded as a genuinely asymptotic distribution.}

\section{Circuits with fixed noise rate in time and xeb transition}\label{sec:strong_noise}

{We shall now consider a more realistic scenario in which the noise strength $\noisestrength$ does not scale inversely to the circuit depth $t$, but still decreases with the system size $N$. This corresponds to setting $\noisestrength = O(N^{-1})$, so that the total number of errors scales linearly with time, i.e.\ $\eta = O(t)$. Specifically, we define 
\begin{equation}
    \eta = \lambda t,
\end{equation}
with fixed $\lambda$, related to the global fidelity, as before, via $F = e^{ -\lambda t}$}. All our results remain applicable in the regime $\lambda t \ll N$, namely whenever the total number of errors remain subleading. Expanding Eq.~\eqref{eq:xeb_sl} for the XEB in the small-$x$ limit, retaining only the first-order term in $x$, substituting $\eta \to \lambda t$, and using $x = e^{-t/\tau} \cdot N/\Nth_0 $, we obtain

\begin{equation}\label{eq:xeb_sl_new}
    \xeb = e^{- \lambda t} + 2  \frac{e^{- t/\tau}}{ \lambda t} \frac{N}{\Nth_0} + O(x^2) \, .
\end{equation}
This expression can be easily interpreted viewing $\xeb$ as the partition function of a 1D system of size $N$ in permutation space, expressed by (see first line of Eq.~\eqref{eq:xeb_sl}): $\xeb = (1| \big(\exp(\frac{x}{N} A) \exp(- \frac{1}{2}\frac{\lambda t}{N} \PP) \big)^N |1) - 1$. As the $\xeb$ is a quantity expressed only in terms of two replicas, the permutation space is limited to identity $\idp$ and swap $\swp$. The term $\exp(- \frac{1}{2}\frac{\lambda t}{N} \PP)$ weights each site with a factor $1$ for identity and $\exp(-\frac{\lambda t}{N})$ for the swap. Each domain wall between the two permutation carry a factor $x/N=\Nth^{-1}(t) =\Nth_0^{-1}e^{-t/\tau}$, therefore at large times (i.e., for small $x$), domain walls between different permutations are suppressed. In this limit, the leading contributions come from configurations where the permutation is uniform across the system. There are two such trivial contributions. The configuration in which the permutation is $\idp$ is exactly canceled by the $-1$ in the definition of $\xeb$. The configuration where the permutation is $\swp$, or pictorially 
\begin{equation}
\begin{tikzpicture}[baseline=(current  bounding  box.center),scale=0.7]
\definecolor{mycolor2}{rgb}{0.0, 0.83, 0.39}
\definecolor{mycolor1}{rgb}{0.83, 0.0, 0.39}
\pgfmathsetmacro{\ll}{2.5}
\draw[line width=0.8mm, mycolor1] (-3*\ll/2,0) -- (-\ll/2,0.);
\node[scale=1.] at (-\ll,-0.35) {$\swp$};
\end{tikzpicture}  \, \, \, ,
\end{equation}
yields $\left(\exp(-\frac{\lambda t}{N})\right)^N=e^{-\lambda t}$, i.e.\ the first term in Eq.~\eqref{eq:xeb_sl_new}. The second term in Eq.~\eqref{eq:xeb_sl_new} (of order $x$) corresponds to the configurations with a single domain wall, pinned either on the left or on the right boundary of the system, 
\begin{equation}
\begin{tikzpicture}[baseline=(current  bounding  box.center),scale=0.7]
\definecolor{mycolor2}{rgb}{0.0, 0.83, 0.39}
\definecolor{mycolor1}{rgb}{0.83, 0.0, 0.39}
\pgfmathsetmacro{\ll}{3.5}
\draw[line width=0.8mm, mycolor2, ultra thick] (0.,0.) -- (2*\ll/3,0.);
\draw[line width=0.8mm, mycolor1] (2*\ll/3,0.) -- (+\ll,0.);

\draw[line width=0.8mm, mycolor1] (3*\ll/2,0.) -- (3*\ll/2+1*\ll/3,0.);
\draw[line width=0.8mm, mycolor2, ultra thick] (3*\ll/2+1*\ll/3,0.) -- (5*\ll/2,0.);
\node[scale=1.] at (\ll/3,-0.35) {$\idp$};
\node[scale=1.] at (\ll*5/6,-0.35) {$\swp$};
\node[scale=1.] at (14*\ll/6,-0.35) {$\idp$};
\node[scale=1.] at (\ll*10/6,-0.35) {$\swp$};
\end{tikzpicture}.
\end{equation}
In fact, each of these two terms contributes to the $\xeb$ as 
\begin{equation}
   \frac{1}{ \Nth_0} \sum_{\ell=1}^N \left(\exp(-\frac{\lambda t}{N})\right)^{\ell}=\frac{N}{\lambda t\Nth_0}  ,
\end{equation}
where $\ell$ denotes the length of the region occupied by the $\swp$ permutation.  A direct inspection of Eq.~\eqref{eq:xeb_sl_new} shows that $\xeb(t)$ is the sum of two exponentials, decaying in time with distinct rates. 
For noise rates $\lambda< \lambda_c = 1/\tau$, the first contribution dominates and coincides with the global fidelity $F(t)=e^{-\lambda t}$. 
In this regime, the late-time behavior of $\xeb$ provides direct access to the fidelity of the \emph{global} circuit. Conversely, for $\lambda>1/\tau$ the second exponential takes over, yielding a decay governed by the noise-independent rate $1/\tau$. \new{This is precisely the transition reported in Refs.~\cite{ware2023sharpphasetransitionlinear,Morvan2023}, where the sharp crossover at $\lambda=\lambda_c$ is interpreted as the breakdown of $\xeb$ as a faithful proxy for fidelity. 
Moreover, it coincides with a transition in the spatial structure of correlations: from an extended, system-spanning regime to a local regime effectively pinned to the boundaries. 
In the latter phase, the dynamics can be viewed as fragmenting the system into mutually incoherent patches, which can be efficiently reproduced by classical patch-based simulations~\cite{Morvan2023}.}

\begin{figure*}[ht!]
    \centering
    \includegraphics[width=\linewidth]{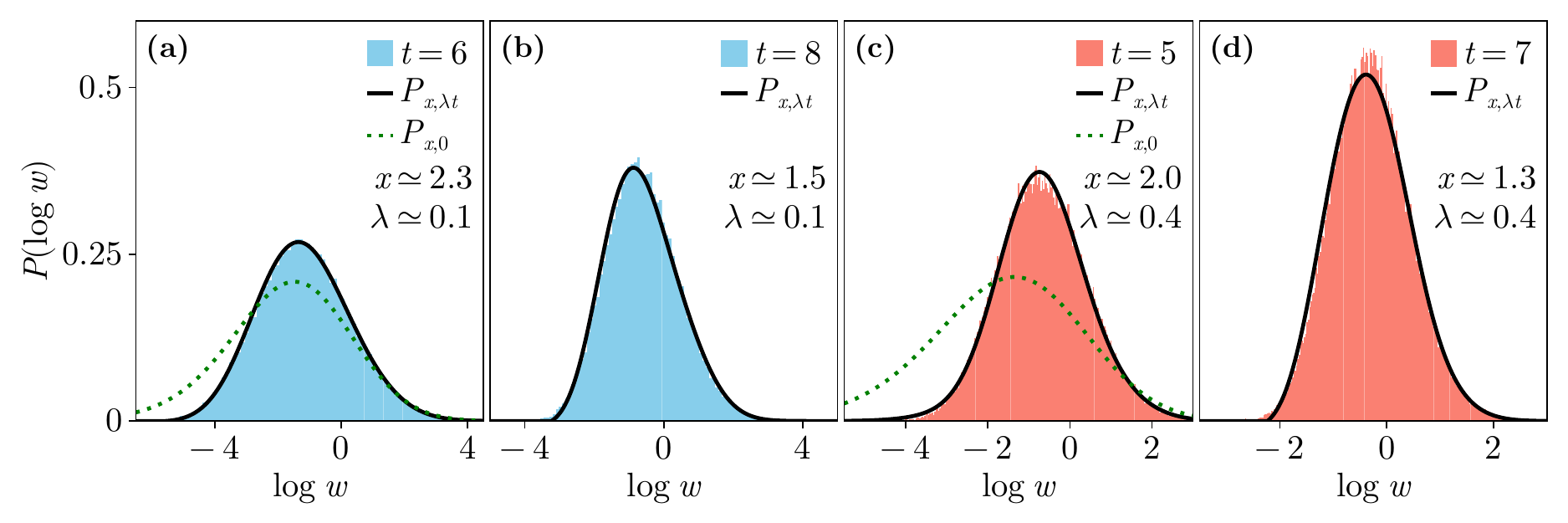}
    \caption{\new{Plot of the PoP for a Haar-random brickwall circuit with $N=32$, affected by \textit{both} two-qubit depolarizing noise (strength $\noisestrength$) and single-qubit amplitude-damping noise (strength $\gamma'$). \new{For each set of parameters, the histogram is built from 10 independent circuit realizations, yielding a total of $10^5$ samples per histogram.} We show two noise settings: $\noisestrength=0.00313$ and $\noisestrength'=0.00442$ in panels (a),(b), which yield a fitted $\lambda\simeq 0.1$, and $\noisestrength=0.016$ and $\noisestrength'=0.022$ in panels (c),(d), which yield $\lambda\simeq0.4$. These correspond to two distinct XEB regimes (see Eq.~\eqref{eq:xeb_sl_new}): $\lambda<1/\tau$ for (a),(b) and $\lambda>1/\tau$ for (c),(d). For each setting we consider two circuit depths $t$, and report the associated rescaled inverse depth $x$ in the plots. The parameters $x$ and $\lambda$ are extracted via the XEB-based fitting procedure described in Sec.~\ref{sec:strong_noise}.
     }}
    \label{fig:circuitdistribution-t}
\end{figure*}

\new{We clarify three points. 
First, we stress that the critical value of $\lambda$ is determined precisely by the scaling of the half-system purity, $\lambda_c = 1/\tau$. Second, \textit{the $\xeb$ at late times \emph{does} provide information about the fidelity $F$ even in the regime $\lambda > 1/\tau$}. In practice, it suffices to determine $\lambda$ by fitting $\lambda$, $\tau$, and $\Nth_0$ simultaneously using the second term of Eq.~\eqref{eq:xeb_sl_new}, and then to compute the fidelity in the usual way as $F = e^{-\lambda t}$. (If available, $\tau$ may alternatively be extracted independently from the scaling of the purity, which can further stabilize the fit). Finally, an analogous transition occurs for all moments $k$ of the distribution~\eqref{eq:TM_sl}. In fact, in complete analogy with the discussion above for the $\xeb$, the small $x$ expansion of $I_k$ yields
\begin{align}
\begin{split}
    &I_k = \\ &= D^{1-k}\left[1 + \frac{k(k-1)}{2}\left( e^{-2 \lambda t}
    + \frac{e^{-t/\tau}}{\lambda t} \frac{N}{\Nth_0}\right)
    + O(x^2,e^{-3 \lambda t})\right] \, ,
\end{split}
\end{align}
so that the approach to the classical distribution $P_{\text{C}}(w) = \delta(w - 1)$ is characterized by either lognormal corrections with parameter $e^{-2\lambda t}$ or $e^{-t/\tau}$. This transition occurs at a critical noise value $ \lambda^{(k)}_{c} \,=\, {1}/{(2\tau)}$, i.e., at half the noise strength associated with the $\xeb$ transition.}

\vspace{5mm}

\begin{figure}[t!]
    \centering \includegraphics[width=\columnwidth]{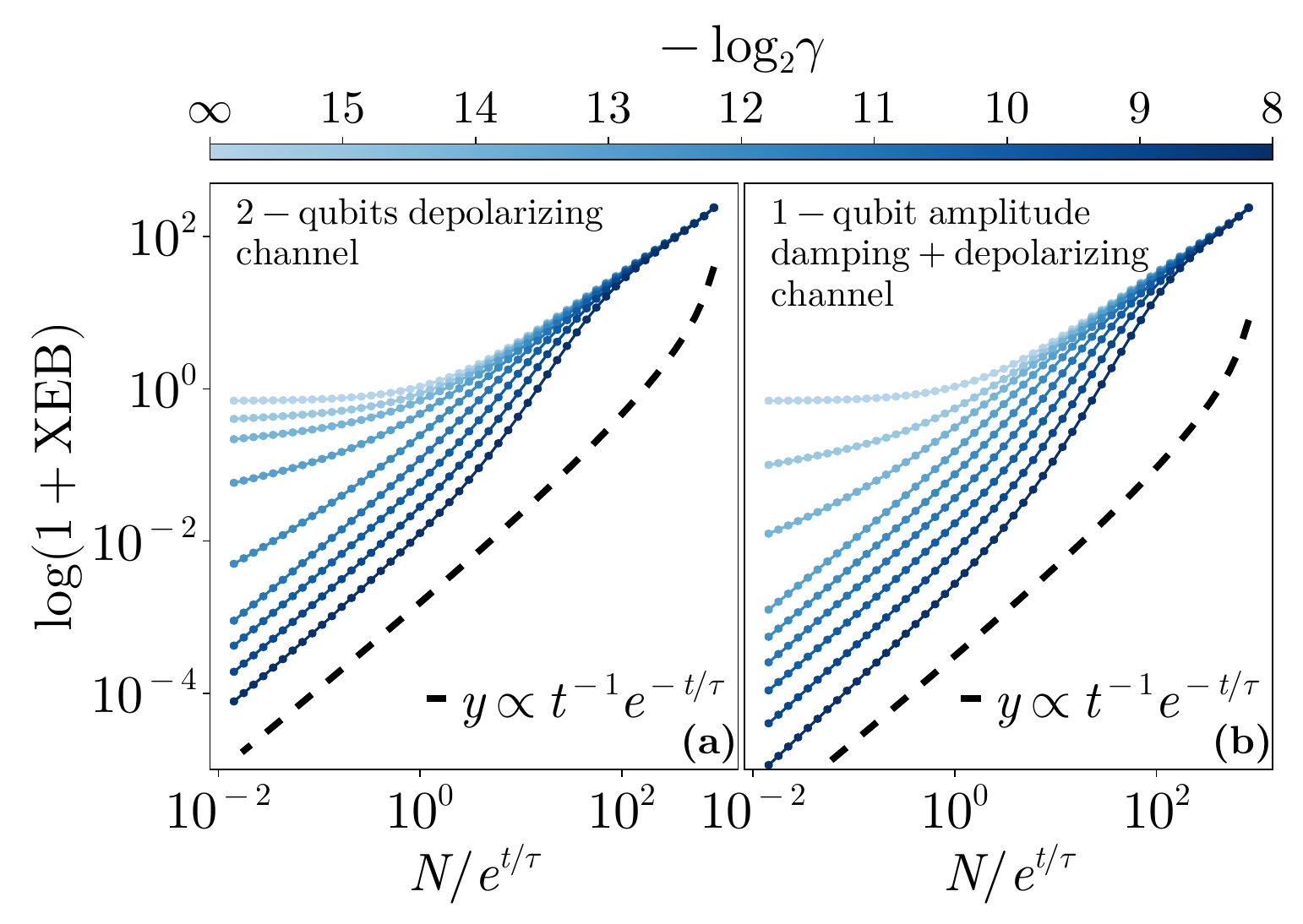}
    \caption{Plot of the $\xeb$ for the brickwall circuit with (a) two-qubits depolarizing (b) single-qubit amplitude damping and depolarizing channels, with different $\noisestrength $, and $N=1024$. We observe the transition at large times in the scaling of $\log(1+\xeb)$, from $-\lambda t$ to $-t/\tau$ as the noise strength $\noisestrength N$ is increased. The crossing happens at $\lambda=1/\tau$ as predicted by Eq.~\eqref{eq:xeb_sl_new} ($\tau= \tau_{\text{bw}}(2)$ in this model). In the regime $\lambda> 1/\tau$, while the slope of the decay is fixed by the noise-independent $\tau$, a shift proportional to $1/\lambda$ as predicted by Eq.~\eqref{eq:xeb_sl_new} is clearly visible, the latter allowing to determine the global fidelity.}   \label{fig:XEB_noscaling}
\end{figure}

As done before in Sec.~\ref{sec:pop}, we again use the $\xeb$ at different values of depth $t$ to determine $x$ and $\lambda$ to reconstruct the PoP $P_{x,\lambda t}(w)$, see Fig.~\ref{fig:circuitdistribution-t}, and with the XEB data shown in Fig.~\ref{fig:XEB_noscaling}. Specifically, we first fit the numerically (experimentally) obtained values $\xeb$ at large times $t$ using Eq.~\eqref{eq:xeb_sl_new}, which allows us to extract $\lambda$, $\tau$, and $\Nth_0$. While one could then directly compute $x(t)$ by assuming the form $x = \frac{N}{\Nth_0} e^{-t/\tau}$, using the fitted values of $\Nth_0$ and $\tau$, this expression may have extra corrections at early times. In practice, we find improved accuracy by  instead adopting the approach used in Sec.~\ref{sec:pop}: after having found $\lambda, \tau, L_0$ from the fit, and then invert Eq.~\eqref{eq:xeb_sl} to obtain $x = x(t, \xeb, \lambda)$. With both $x$ and $\lambda$ determined in this way, we reconstruct the analytical distribution $P_{x,\lambda t}(w)$ with moments given by Eq.~\eqref{eq:TM_sl}. \new{We stress that the fact that the values of $x$ and $\eta$ obtained from the $\xeb$ are also the ones fixing the distribution $P_{x,\eta}$ is a non-trivial check of the universality and the validity of Eq.~\eqref{eq:xeb_sl} and Eq.~\eqref{eq:TM_sl} for all moments $k$. Moreover, our results extend beyond the specific setting considered here to include general chaotic circuits and various forms of noise. In fact, Appendix~\ref{ref:additionalnumerical} provides an example based on a \emph{deterministic gate set}, thus going beyond the paradigm of random quantum circuits. Specifically, in Fig.~\ref{fig:distribution_floquet} we consider the chaotic kicked Ising model with dephasing noise and observe good agreement with our predictions.}


\section{Conclusions}
In this work we have investigated the universal properties of anticoncentration in generic, chaotic 1D quantum many‐body systems subject to weak noise. Whereas the influence of noise on infinite-depth circuits, characterized by the Porter–Thomas distribution of overlaps, has been recently studied, see for example \cite{Shaw_2024}, here we have focused on \textit{the interplay between finite circuit depth and noise}, a question of central relevance for near-term quantum devices. We derived a universal form of the distribution of bit-strings probabilities (PoP distribution), as a function of two dimensionless parameters $x$ and $\eta$ determined by circuit depth, system size, and the global fidelity. As we have shown, these parameters can be extracted in an experimental setting by evaluating the linear cross entropy $\xeb$ across different circuit depths, for any values of noise strength, at least in the regime where the number of error remains subleading $- \log F \ll N$. 
Given that our interest is in shallow circuits, this regime is achieved even for very strong noise strengths at small depths, and indeed our universal PoP distribution provides an excellent fit of our numerical data for quite shallow circuits and strong noise. 

Our analysis is based on an exact mapping to a one-dimensional model of random matrix product operator, that are generated by the action of a generic noisy staircase circuit. Beyond serving as a tractable theoretical toy model, noisy RMPO are themselves experimentally accessible on current quantum platforms~\cite{piroli2021quantum,malz2024preparation,smith2024kevin,david2024preparing,zhang2024yifan}, whereas random MPO can also be used to benchmark of quantum experiments \cite{Piroli2025}. Thus, our results not only furnish clear predictions for state‐of‐the‐art quantum machines that can be directly used for useful benchmarking, but also clarify the persistence of genuine quantum effects in the presence of realistic noise. 

While the universality of our results encompasses numerous realistic systems,  it may also be interesting to explore PoP shapes beyond this universality class, which requires a certain amount of fine-tuning: dual-unitary circuits~\cite{Claeys2025}, the presence of conserved charges \cite{PhysRevLett.123.210603, PhysRevLett.131.210402,PhysRevX.13.011045}, integrability would have implications on our results (even in the absence of noise)~\cite{PhysRevX.14.031048}. We leave these settings to future works.

\section{Acknowledgments}

We acknowledge discussions with Xhek Turkeshi
and for collaborations on related topics. J.D.N. and G.L. are funded by the ERC Starting Grant 101042293 (HEPIQ) and the ANR-22-CPJ1-0021-01. ADL acknowledges support by the ANR JCJC grant ANR-21-CE47-0003 (TamEnt). This work was granted access to the HPC resources of IDRIS under the allocation AD010914149R2 and AD010613967R2. 

\paragraph*{Code and Data Availability.} The code and the data for our simulations will be publicly shared at publication.

\newpage 

\appendix

\section{Derivation of the transfer matrix for RMPO}\label{sec:app_singlequdit}

Here, we provide a full derivation of the replica tensor network for the noisy RMPO model giving Eq.~\eqref{eq:contraction}. In Eq.~\eqref{eq:mps_circuit_2}, unitaries $U_i$ \new{and $V_i$}, of size $d\chi \times d\chi$, act on $r + 1$ qudits, with $r=\log_d \chi$. 
 To compute the overlap $w_{\boldsymbol{x}}=D \llangle \boldsymbol{x}, \boldsymbol{x}|\rho_{\Noise}(U) \rrangle$ for a fixed bit-string $\boldsymbol{x}$, one must contract the open physical legs in Eq.~\eqref{eq:mps_circuit_2} with the state $\llangle \boldsymbol{x},\boldsymbol{x}|$; that is,  
 \begin{equation}
 \begin{split}
 w_{\boldsymbol{x}} &=D\llangle \boldsymbol{x},\boldsymbol{x}|\rho_{\Noise}(U) \rrangle 
 = \new{ D \llangle \boldsymbol{x},\boldsymbol{x}|  \prod_{i=1}^{N-r} W_i|\boldsymbol{0},\boldsymbol{0}\rrangle\, },
\end{split}
\end{equation}
\new{where we defined 
\begin{equation}
    W_i = \big( (V_i \otimes V_i^*)\cdot\Noise_i (U_i \otimes U_i^*) \big) 
\end{equation}
}
and where we omit dependence on the noise strength $\noisestrength$ for readability. The IPR $I_k = D^{-k} \sum_{\boldsymbol{x}} \mathbb{E}[w_{\boldsymbol{x}}^k]$ is then obtained by replicating the network $k$ times, summing over all $\boldsymbol{x}$ and performing the average over the $U_i$ \new{and $V_i$}. This results in
 \begin{equation}
 \label{eq:IPRmodel}
 \new{I_k =  \sum_{\boldsymbol{x}} \left( \llangle \boldsymbol{x},\boldsymbol{x}|  \prod_{i=1}^{N-r} \Ex[W_i] |\boldsymbol{0},\boldsymbol{0}\rrangle\right)^{\otimes k} ]\, }.
 \end{equation}
 Now we employ the modified Weingarten formula (Eq.~\eqref{eq:noisy_wg_new_1})
 \new{\begin{align}
 \label{eq:weingchid}
     \begin{split}
       \Ex[W_i]&=\mathbb{E}_{U,V \sim \Haar(d\chi)} \left[ \left((V_i \otimes V_i^*)\cdot\Noise_i (U_i \otimes U_i^*)\right)^{\otimes k} \right] \\ &=\sum_{\pi, \sigma \in S_k} \tilde{\Wg}_{\pi, \sigma}(d \chi,\noisestrength)\, |\sigma\rangle\!\rangle_d |\sigma\rangle\!\rangle_{\chi}  \langle\!\langle \pi|_{\chi} \langle\!\langle \pi|_{d}  \, , 
     \end{split}
 \end{align}}
 where we explicitly used the factorization \new{of permutation states over distinct Hilbert spaces} $|\sigma\rangle\!\rangle_{d \chi} =|\sigma\rangle\!\rangle_d |\sigma\rangle\!\rangle_{\chi}$. This equation has to be applied for \new{each $W_i$, i.e.\ for each pair $U_i$ and $V_i$ of matrices}. \new{In the $d-$dimensional spaces, permutations are either contracted with the initial state $|0,0\rangle$, or with the final state $|x_i,x_i \rrangle$. Since these states are normalized to one, the result is an irrelevant factor: $( \llangle 0,0 |) \sigma \rrangle_d^{\otimes k} = \llangle \pi| ( |x_i,x_i \rrangle)_{d}^{\otimes k} = 1$. The sum over $x_i$ simply yields a multiplicative factor of $d$ for each qubit, i.e.\ an overall contribution $D=d^N$. Because of these free boundary conditions, the left and right boundary vectors can be written simply as $(L|=(\pmb{1}|$ and $|R)=\Tilde{\Wg}(d\chi,\noisestrength)|\pmb{1})$. The Weingarten matrix in the right boundary vector comes from the average of the last gate of the staircase. In the bulk, permutations resulting from consecutive gates 
 $V_i, U_{i+1}$ are contracted together, giving rise to the overlaps matrix $G_{\sigma,\pi}(\chi)= \llangle\sigma|\pi\rrangle_\chi$. Putting together all these terms one finally obtain Eq.~\eqref{eq:contraction}, with the transfer matrix defined as in Eq.~\eqref{eq:TM}. Notice that in Eq.~\eqref{eq:contraction} there are in total $N-r$ Weingarten matrices (one incorporated in $|R)$ and $N-r-1$ in the term $T^{N-r-1}$), as should be as the staircase circuit Eq.~\eqref{eq:mps_circuit_2} features exactly $N-r$ unitaries.  }

The derivation of the $\xeb$ for this model follows similarly, now involving only $k=2$ replicas: one noisy and one noiseless. \new{The overlaps matrix remains unchanged, while the Weingarten matrix is now defined by
\begin{align}\label{eq:noisy_wg_new_xeb}
    \begin{split}
& \Ex_{U,V}[ \left((V \otimes V^{*})\cdot\Noise_{\noisestrength}\cdot(U \otimes U^{*})\right)\otimes \left((V \otimes V^{*})\cdot(U \otimes U^{*})\right)] \\ &=\sum_{\pi, \sigma \in S_2} \left[\tilde{\Wg}_\xeb\right]_{\pi, \sigma}(q,\noisestrength)\, |\sigma\rangle\!\rangle \langle\!\langle \pi| \, ,
    \end{split}
\end{align}
and can be computed as before using $\tilde{\Wg}_\xeb = \Wg \, \tilde{G}_\xeb \, \Wg$, and defining
\begin{equation}
[\tilde{G}_\xeb]_{\sigma',\sigma}(q,\noisestrength) = \llangle \sigma' \mid \Noise_{\noisestrength} \otimes \mathbb{1} \mid \sigma \rrangle_q \, ,
 \end{equation}
with $\sigma',\sigma \in S_2$. Calculations analogous to those in Sec.~\ref{subsec:Generic noise} then yield
\begin{equation}
   \tilde{G}_\xeb(q,\noisestrength)=q^2\begin{pmatrix}
     1 & 1/q \\ 1/q & 1-\noisestrength  
      \end{pmatrix} +O(\noisestrength^2)  \ , 
\end{equation}
where again we assumed $\kappa=1$. This equation is the analogous of Eq.~\eqref{eq:G_tilde_final}. We can now obtain the expansion at first order in $q^{-1}$ and $\noisestrength$ for the noisy Weingarten matrix 
\begin{equation}
    \tilde{\Wg}_\xeb(q,\gamma) = q^{-2}\left(\begin{pmatrix}
        1 & -\frac{1}{q} \\  -\frac{1}{q} & 1-\noisestrength
    \end{pmatrix}+O(\gamma^2,q^{-2},\gamma q^{-1})\right)
\end{equation}}

\new{As for $I_2$, we can write the replica contraction  $\xeb=D(L|T_\xeb^{N-r-1}|R)-1$ with $T_\xeb=\tilde{\Wg}_\xeb(d\chi,\gamma)G(\chi)$, $(L|=(\pmb{1}|$ and $|R)=\tilde{\Wg}_\xeb(d\chi,\noisestrength)|\pmb{1})$. Scaling the noise strength as previously done $\noisestrength=\eta/N$, we get }
\begin{equation}
   T_\xeb = d^{-1} \left( \Id + \frac{x}{N} A - \frac{\eta}{2N} \PP + O(N^{-2}) \right) \ , 
\end{equation}
which, in the scaling limit and after exponentiation, leads to  Eq.~\eqref{eq:xeb_sl}.\\

The calculation of the average state fidelity $F=\Ex_U[\mathrm{Tr}(\rho_\Noise(U)\rho(U))]$ is very similar to the one of $\xeb$. The trace in the fidelity expression can be evaluated using the ‘swap trick’ by rewriting $\mathrm{Tr}(\rho_{\Noise} \rho)=\mathrm{Tr}((\rho_{\Noise} \otimes \rho)\swp)=\llangle \rho_{\Noise} \otimes \rho|\swp\rrangle$, where $\swp$ denotes here the swap operator (or equivalently the transposition permutation). \new{This adds boundary terms in the replica contraction, which we encapsulates inside a matrix $[\Lambda_F]_{\sigma' \sigma}(d) = \delta_{\sigma' \sigma} \llangle s | \sigma \rrangle_d=\mathrm{diag}(d, d^2)$, so that the replica contraction now reads $F=(L|T_F^{N-r-1}|R)$ with $T_F=\tilde{\Wg}_\xeb(d\chi,\noisestrength)\Lambda_F(d)G(\chi)$, $(L|=(\pmb{1}|$ and $|R)=\tilde{\Wg}_\xeb(d\chi,\noisestrength)\Lambda_F(d\chi)|\pmb{1})$. In the scaling limit, we notice that the right boundary vector becomes simply $|R)=(0,1)$.
Finally, after replacing $\noisestrength=\eta/N$, we get 
\begin{equation}
T_F\scaleq\begin{pmatrix}
    1/d & 0\\ (d+1)x/(dN) & 1-\eta/N
\end{pmatrix} \, ,
\end{equation}}
which in the scaling limit yields $F\scaleq e^{-\eta}$.

\section{Random phase model}
\label{sec:RPM}
To begin, let us briefly revisit the standard definition of the Random Phase Model (RPM)~\cite{ChanDeLucaChalker2018PRL, Christopoulos_2025}. 
This model consists of a random circuit acting on $N$ qudits and defined the subsequent action of two different layers at time step $t$: first, on each site $i$, one acts with single-site Haar-random unitary, $U_{i}^{(1)}(t)$; then, on each bond between neighboring sites $(j, j+1)$, one acts with a diagonal two-site gate whose entries are phases $[U_{j,j+1}^{(2)}(t')]_{x_j x_j+1, x_j x_j+1}=\exp[i \varphi^{(j)}_{x_j,x_{j+1}}(t')]$ ($x_j, x_{j+1} \in \{ 0,1\ldots, d-1 \}$). Each phase $\varphi_{x_j,x_{j+1}}^{(j)}(t')$ is an independent Gaussian random real variable with mean zero and variance $\epsilonRPM$, which controls the coupling strength between neighboring spins. A given realization of the circuit is obtained by drawing independently for each $j$ and $t'$ the single-site unitaries and the phases $\varphi^{(j)}_{x_j,x_{j+1}}(t')]$. 
Consider the $k$-th moment of the output distribution $w^k = D\langle\boldsymbol{x} | W_{\rm RPM}(t) |\boldsymbol{x}\rangle$, where $W_{\rm RPM}(t)$ denotes the global RPM circuit with $t$ repeated action of single-site and two-site layers. As explained in \cite{Christopoulos_2025}, the circuit average $\mathbb{E}[w^k]$ in the absence of noise can be computed at large $d$ in two steps: one first averages over the single-site unitaries using once again the Weingarten formula
\begin{multline}
\label{eq:weingd}
    \mathbb{E}_{\rm RPM}[(U^{(1)}_i(t') \otimes U^{(1)}_i(t')^\ast)^{\otimes k}] =\\= \sum_{\pi, \sigma \in S_k} \Wg_{\pi, \sigma}(d)\, |\sigma\rangle\!\rangle_d  \langle\!\langle \pi|_{d} = 
     d^{-k} \sum_{\sigma \in S_k} \, |\sigma\rangle\!\rangle_d  \langle\!\langle \sigma|_{d}
\end{multline}
Note that this equation is formally analogous to~\eqref{eq:weingarten}, but it is restricted to the action on the $d$-dimensional local physical Hilbert space. Beyond allowing the expansion of Weingarten functions, the large $d$ limit also implies that the same permutation $\sigma_i \in S_k$ at a given site $i$ propagates at all times, as the overlaps $G_{\sigma',\sigma}(d)=\llangle \sigma'|\sigma\rrangle_{d}  \sim d^k \delta_{\sigma',\sigma} $. Thus, the effect of the $2$-site gates can be accounted for computing
\begin{equation}
    \llangle \sigma_j, \sigma_{j+1} | U_{j,j+1}^{(2)}(t') | \sigma_j \sigma_{j+1}\rrangle =  d^{2k} e^{-\epsilonRPM (k - \mathrm{n_F}(\sigma_j,\sigma_{j+1}))} \, ,
\end{equation}
where $\mathrm{n_F}(\sigma_j,\sigma_{j+1})$ denotes the number of common fixed points between $\sigma_j$ and $\sigma_{j+1}$.
Collecting the contribution from the $t$ layers of phases, we have that
\begin{equation}
    \mathbb{E}[w^k] = \sum_{\sigma,\sigma'} [T_{\rm RPM}^N]_{\sigma, \sigma'} \;, \quad   [T_{\rm RPM}]_{\sigma', \sigma} = e^{-\epsilonRPM t (k - \mathrm{n_F}(\sigma,\sigma'))}
\end{equation}
In order to include the effect of noise, we assume for simplicity that it affects each single-site Haar unitary, with the presence of a depolarizing channel $\Noise_{\noisestrength}^{\text{dep.}}$. Consequently, as in Eq.~\eqref{eq:noisy_wg}, we replace $\Wg(d)$ with $\tilde{\Wg}(d, \noisestrength)$ in \eqref{eq:weingd}. For large $d$, Eq.~\eqref{eq:noise_wg_exp} yields
\begin{equation}
    \tilde\Wg(d, \noisestrength) = d^{-k}\left(1 - \noisestrength \PP + O(\noisestrength^2) + O(d^{-1}) \right) \, .
\end{equation}
We thus arrive at 
\begin{equation}
\label{eq:wRPMnoisy}
    \mathbb{E}[w^k] = \sum_{\sigma,\sigma'} [\tilde T_{\rm RPM}^N]_{\sigma, \sigma'} \;, \quad \tilde T_{\rm RPM} = T_{\rm RPM} e^{-\PP \noisestrength t + O(\noisestrength^2)} 
\end{equation}
Now, we recall that for the RPM, the Thouless length can be identified with $\mathsf{L}(t) = e^{2\epsilonRPM t}$ and at large time, one can rewrite
\begin{equation}
    T_{\rm RPM}  = 1 + \frac{A}{\mathsf{L}(t)} + O(\mathsf{L}(t)^{-2}) \sim e^{A/\mathsf{L}(t)} \, .
\end{equation}
Finally, in the scaling limit \eqref{eq:def_x} and $ \noisestrength=\eta/(N t)$, we can combine the two factors in $\tilde T_{\rm RPM}^N \sim [e^{A x/N} e^{-\PP \eta/N}]^N \to e^{A x - \PP \eta}$, in agreement with \eqref{eq:TM_sl} in the main text. 

\section{Additional numerical data}\label{ref:additionalnumerical}

\begin{figure}[H]
    \centering \includegraphics[width=\columnwidth]{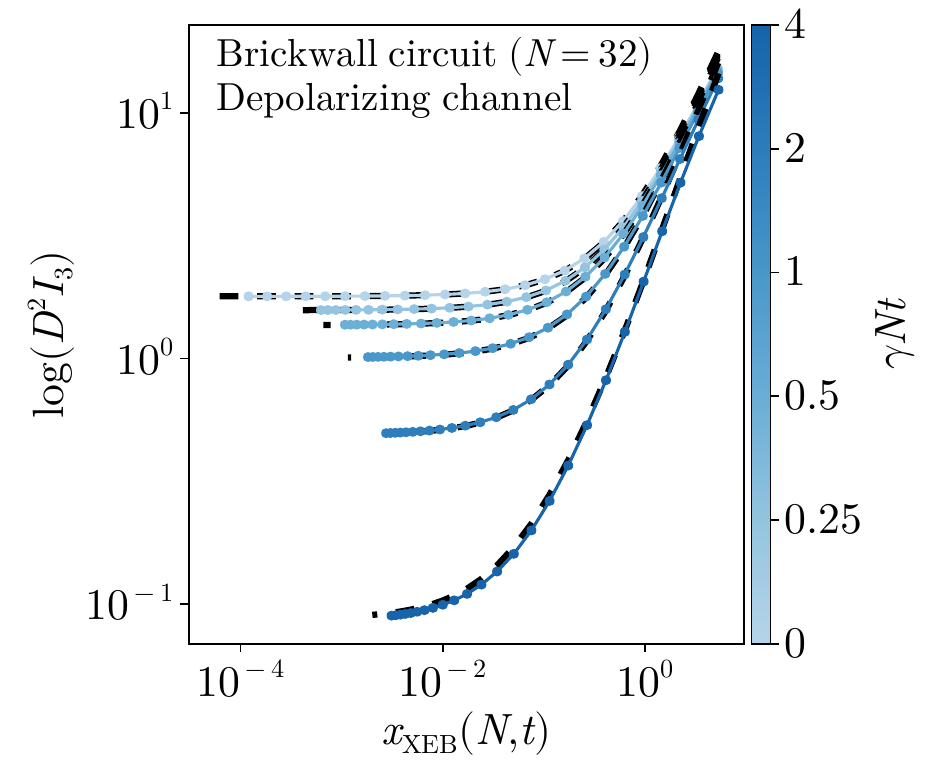}
    \caption{Plot of $I_3$ for the brickwall circuit with two-qubits depolarizing channel with noise strength $\noisestrength \propto 1/(Nt)$, with $N=32$. We show it as a function of $x_\xeb(N,t)$ which is obtained by computing first the $\xeb$ at large time to estimate $\eta$ and then by computing the $\xeb$ at time $t$, and inverting Eq.~\eqref{eq:xeb_sl} to obtain $x$. We show also the analytical prediction Eq.~\eqref{eq:TM_sl} (black dashed line).}   \label{fig:I3}
\end{figure}

In Fig.~\ref{fig:I3}, we show $I_3$ for the brickwall circuit. The plot is analogous to the plot of $\xeb$ in Fig.~\ref{fig:XEBplot}. We show that the $x$ extracted from the $\xeb$, with the procedure detailed in the main text, {gives good predictions beyond $2$-replicas observables through Eq.~\eqref{eq:TM_sl}. This provides an additional check that all the moments of the anticoncentration, namely the whole PoP distribution, depend only on the two fitting parameters $\eta$ and $x$.}\\

\begin{figure}[t!]
    \centering \includegraphics[width=\columnwidth]{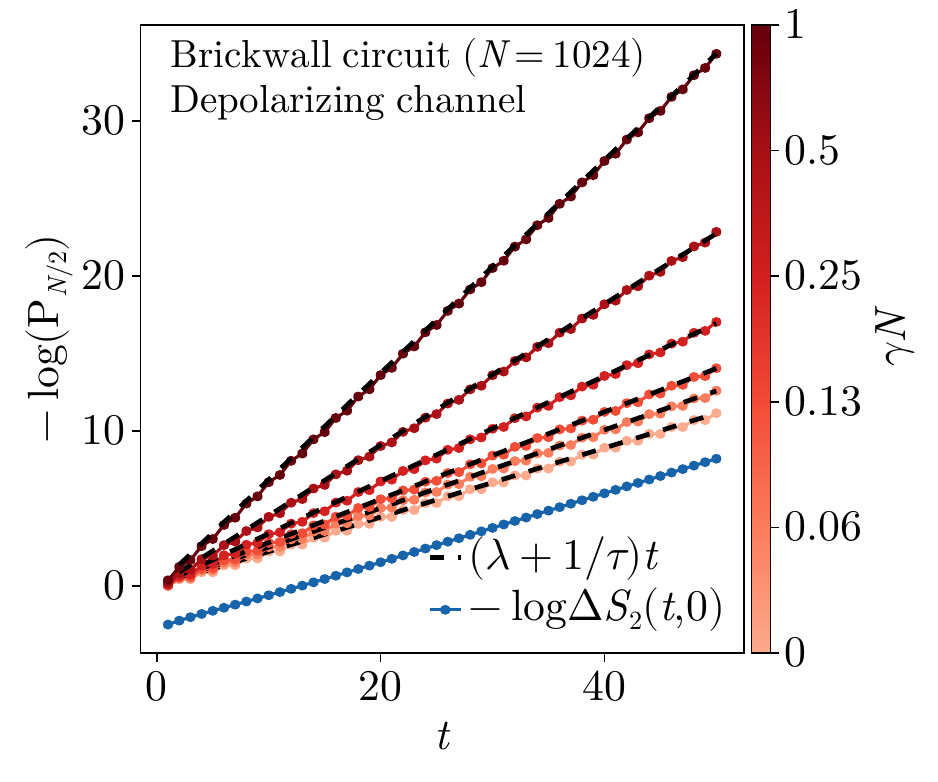}
    \caption{Scaling of the half-chain purity for the brickwall circuit with two-qubits depolarizing channel and noise strength $\noisestrength\propto 1/N$. The system size is $N=1024$. We show that this scaling is $\exp(-t\cdot(\lambda+1/\tau))$, where the parameter $\lambda$ is obtained through a fit from the $\xeb$ at large times. $\tau$, given by Eq.~\eqref{eq:tau}, 
    is the scaling of the purity in the noiseless case $\lambda=0$.
    It also corresponds to the scaling of $\Delta S_2(t,0)=S_2(\infty,0)-S_2(t,0)$ (see blue line), where $S_2(t,\lambda)=-\log I_2(t,\lambda)$ is the participation entropy and $S_2(\infty,0)=-\log I_2^{\text{PT}}$.(For visualization purposes, the blue line is shifted by a constant of $3$).}\label{fig:Purity}
\end{figure}

In Fig.~\ref{fig:Purity}, we show that the average half-chain purity 
\begin{equation}
    \mathrm{P}_{N/2}= \Ex[\mathrm{Tr}(\rho_\mathrm{N/2}^2)] \, ,
\end{equation}
with $\rho_{N/2}=\mathrm{Tr}_{1,...,N/2}(\rho_{\Noise}(U))$, scales as $e^{-t(\lambda+1/\tau)}$. This observation comes from the fact that we can write $\mathrm{P_{N/2}}=\mathrm{Tr}\big(\rho^{\otimes 2} (s^{\otimes N/2} \otimes \Id^{\otimes N/2})\big)$, where $s$ is the swap operator. This forces a unique domain configuration 
\begin{equation}
\begin{tikzpicture}[baseline=(current  bounding  box.center),scale=0.7]
\definecolor{mycolor2}{rgb}{0.0, 0.83, 0.39}
\definecolor{mycolor1}{rgb}{0.83, 0.0, 0.39}
\pgfmathsetmacro{\ll}{3.5}
\draw[line width=0.8mm, mycolor1, ultra thick] (0.,0.) -- (\ll/2,0.);
\draw[line width=0.8mm, mycolor2] (\ll/2,0.) -- (+\ll,0.);

\node[scale=1.] at (\ll/6,-0.35) {$\swp$};
\node[scale=1.] at (\ll*5/6,-0.35) {$\idp$};
\end{tikzpicture}.
\end{equation}
The contribution of the identity $e$ is $1$ while the contribution of the swap $s$ is $\lim_{N\to \infty}\left(1-\frac{\eta t}{N}\PP_{ss}\right)^{N/2} =e^{-\eta t}$. The unique domain wall gives $\Nth^{-1}=\Nth_0^{-1}e^{-t/\tau}$. Collecting all these contributions gives the correct scaling.\\

\begin{figure}[t!]
    \centering
    \includegraphics[width=\columnwidth]{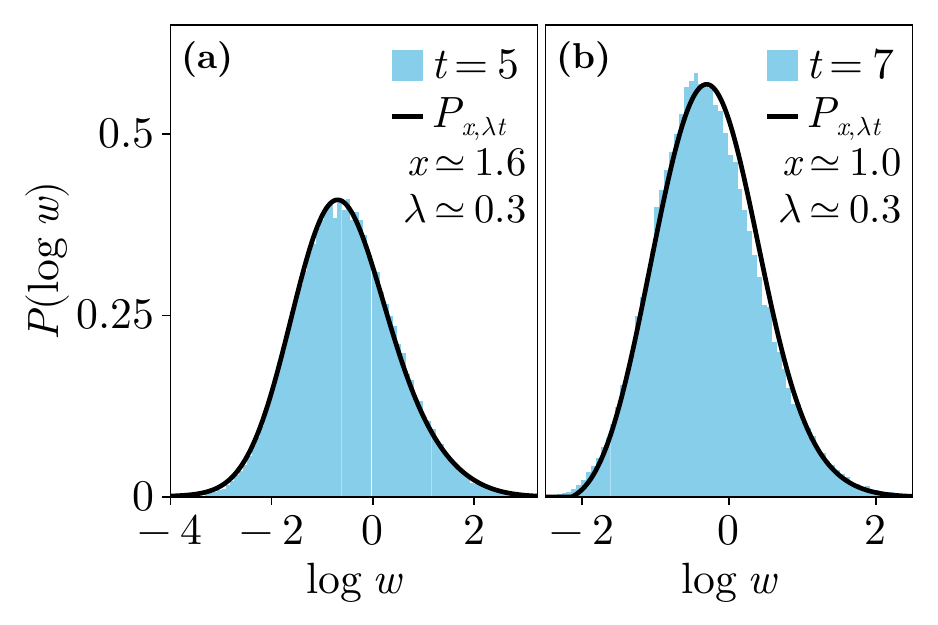}
    \caption{\new{Plot of the PoP for the kicked Ising model with $N=32$ subject to single-qubit dephasing noise of strength $\noisestrength=0.00781$. For each set of parameters, the histogram is built from 10 independent circuit realizations, yielding a total of $10^5$ samples per histogram. We present the distribution at two different times and extract the parameters $x$ and $\lambda$ through a maximum likelihood procedure. The resulting values are shown in the figure. The close agreement between the estimate of $\lambda$ and of the half-chain purity timescale $\tau^{-1}_{\mathrm{Floquet}}$, reveals that the system operates at the crossover between the two $\xeb$ regimes.}}
     \label{fig:distribution_floquet}
\end{figure}

\new{To test the robustness of our universality claims beyond ensembles of random unitaries, we consider the kicked Ising model (KIM)~\cite{Tirrito2024,Prosenfloquet1,Prosenfloquet2,Prosenfloquet3}. The dynamics is generated by the Floquet operator}
\begin{equation}
    \label{eq:Floquet operator}
    U_\mathrm{F}= e^{-i b\sum_j X_j }e^{-i\left(h\sum_j Z_j + J\sum_j Z_j Z_{j+1}\right)}\,,
\end{equation}
\new{applied at each time step. We fix the parameters to be \(J=1\), \(b=(\sqrt{5}+5)/8\), and \(h=(\sqrt{5}+1)/4\). Our conclusions are not expected to depend sensitively on these particular values, provided the model remains in a non-integrable regime. To mitigate basis-dependent effects, the Floquet evolution is initialized from a random product state. We consider as a source of noise a single-qubit quantum dephasing channel}
\begin{equation}
    \Noise(\rho)=(1-\noisestrength)\rho +\noisestrength Z\rho Z \,.
\end{equation}

\new{Fig.~\ref{fig:distribution_floquet} shows the empirical and analytical distribution for this model. Due to the absence of efficient numerical techniques to estimate the $\xeb$ at large time, the two parameters of the analytical distribution are fitted through maximum likelihood estimation. The fitted $\lambda$ can be compared with the half-chain purity timescale $\tau^{-1}_{\mathrm{Floquet}}=0.28$ obtained for the same model in a recent study~\cite{Sauliere_2025}.}

\bibliography{bib}
\bibliographystyle{apsrev4-2}

\end{document}